\documentclass[aps, prd, amsmath, floats, floatfix, twocolumn,superscriptaddress, nofootinbib, showpacs,eqsecnum]{revtex4-1}

\usepackage{graphicx}

\newcommand{\bi}{\bibitem}
\newcommand{\be}{\begin{eqnarray}}
\newcommand{\ee}{\end{eqnarray}}

\newcommand{\eq}{\begin{equation}}
\newcommand{\eeq}{\end{equation}}

\newcommand{\mnras}{Mon.\ Not.\ Roy.\ Astron.\ Soc.\ }
\newcommand{\apjl}{Astrophys.\ J.\ Lett.\ }

\usepackage{array}
\usepackage{ulem}

\begin{document}

\title{The final stages of accretion onto non-Kerr compact objects}

\author{Cosimo Bambi}
\email{cosimo.bambi@ipmu.jp}

\affiliation{
Institute for the Physics and Mathematics of the Universe, 
The University of Tokyo, Kashiwa, Chiba 277-8583, Japan}

\affiliation{
Arnold Sommerfeld Center for Theoretical Physics,
Ludwig-Maximilians-Universit\"at M\"unchen, 80333 Munich, Germany}

\author{Enrico Barausse}
\email{barausse@umd.edu}

\affiliation{
Maryland Center for Fundamental
    Physics \& Joint Space-Science Institute,
  Department of Physics, University of Maryland, College
    Park, MD 20742, USA}

\date{\today}

\preprint{IPMU11-0132}

\begin{abstract}
The $5 - 20 M_\odot$ dark objects in X-ray binary systems and 
the $10^5 - 10^9 M_\odot$ dark objects in galactic nuclei are currently thought to be the Kerr black holes 
predicted by General Relativity. However, direct observational
evidence for this identification is still elusive, and the only viable approach to 
confirm the Kerr black hole hypothesis is to explore and rule out any other possibility. 
Here we investigate the final stages of the accretion process onto 
generic compact objects. While for Kerr black holes and for more 
oblate bodies the accreting gas reaches the innermost stable 
circular orbit (ISCO) and plunges into the compact object, we find 
that for more prolate bodies several scenarios are possible, depending 
on the spacetime geometry. In particular, we find examples in which
the gas reaches the ISCO, but then gets trapped between the ISCO and the compact object. In this situation, accretion 
onto the compact object is possible only if the gas loses additional angular momentum, 
forming torus-like structures inside the ISCO.
\end{abstract}


\maketitle


\section{Introduction}

There is by now robust observational evidence supporting the
existence of $5 - 20 M_\odot$ dark objects in X-ray binary systems 
and that of $10^5 - 10^9 M_\odot$ dark objects in galactic nuclei~\cite{ram}. 
These objects are thought to be the black holes (BHs) predicted by General 
Relativity (GR), since their characteristics cannot be explained otherwise without introducing 
new physics. In 4-dimensional GR, an (uncharged) 
stationary BH is described by the Kerr solution and is uniquely
characterized by two parameters, the mass $M$ and the spin angular 
momentum $J$~\cite{hair}. A fundamental limit for a Kerr BH 
is the bound $|a| \le 1$, where $a = 
cJ/(G M^2)$ is the spin parameter, which ensures that the Kerr metric describes a BH and not
a naked singularity.

However, the evidence that the geometry around these BH 
candidates is described by the Kerr metric is still circumstantial,
and the Kerr BH hypothesis mainly relies on the assumption
that GR is the correct theory of gravity.
Because GR has been tested only in the weak-field regime~\cite{will},
several authors have suggested possible ways to further test the Kerr BH hypothesis using 
future, and in some cases even present, data.

More specifically, the detection of extreme mass 
ratio inspirals (EMRIs, i.e. systems consisting of a stellar-mass
compact object orbiting a supermassive BH candidate)
with future space-based gravitational-wave detectors 
will allow one to map the spacetime geometry
with exquisite accuracy. This is because 
missions like LISA or a similar European mission will
be able to follow the stellar-mass
compact object for millions of orbits around the central supermassive BH candidate,
and therefore deviations from the Kerr geometry will lead to a phase
difference in the gravitational waveforms that grows with the number of observed cycles.
This technique is very promising and has been studied in detail by many authors~\cite{emris,hydrodrag,gair}.
Likewise, future gravitational-wave detectors will
be able to detect the quasi-normal modes of BH
candidates, and because for a Kerr BH the frequencies of these
modes depend only on the spacetime's mass $M$ and spin $J$,
any departure from this pattern will allow one to measure deviations away from the
Kerr geometry~\cite{qnm}.

The geometry around BH candidates can be constrained also
with present or future electromagnetic data.
For instance, radio timing observations allow constraints to be put
on the quadrupole moment of the compact companions of radio pulsars~\cite{wex},
while astrometric monitoring 
of stars orbiting at milliparsec distances from Sgr A$^\star$ may be 
used to constrain models for the supermassive BH candidate 
at the center of the Galaxy~\cite{orbits}.
Constraints on the nature of BH candidates can also be
obtained by extending the methods currently used to estimate the spins of
these objects, such as X-ray continuum~\cite{continuum} and iron-K$\alpha$  measurements~\cite{iron},
observations of quasi-periodic oscillations~\cite{qpos}, and measurements of the cosmic X-ray background~\cite{agn}.
These methods can in principle be applied even with present data, provided that the systematic errors are properly understood. 
Future observations of the shadow of nearby supermassive BH candidates 
are another exciting possibility to test the BH paradigm~\cite{shadow}.

BH candidates are often surrounded by an accretion disk. For
the stellar-mass objects in X-ray binary systems, the disk 
originates from the material stripped from the companion, while
the gas accreting onto the supermassive objects in galactic nuclei 
comes from the interstellar medium. When the mass accretion 
rate is moderate and the accreting gas has significant angular 
momentum, the disk is geometrically thin and optically thick. The 
standard framework to describe these disks is the Novikov-Thorne model~\cite{ntmod}, where the
disk lies on the equatorial plane of the system and the gas moves 
on nearly geodesic circular orbits. If the central object is a BH, 
the inner edge of the disk is assumed to be at the radius of the
innermost stable circular orbit (ISCO): circular orbits inside the
ISCO are radially unstable, so the gas reaches the ISCO moving on quasi-circular orbits,
and then quickly plunges into the BH. 
A crucial ingredient of the Novikov-Thorne model is the assumption 
that the gas does not emit additional radiation as soon as it enters 
the plunging region. Numerical simulations show that deviations 
from this picture are small relative to the effect of the uncertainties
in other parameters of the system~\cite{cfa} and can therefore be neglected.  
(Note however that the authors of Ref.~\cite{krolik}
reach a different conclusion, because using general relativistic magneto-hydrodynamics simulations
they find that there can be significant magnetic stress inside the ISCO.)

If the central object is not a Kerr BH, 
the final stages of the accretion process may be different. 
In this work we focus on the so-called Manko-Novikov (MN) spacetimes~\cite{mn}.
These are stationary, axisymmetric, and asymptotically flat
exact solutions of the vacuum Einstein equations with arbitrary mass-multipole moments, 
and they can therefore describe
the geometry outside a generic compact object, be it a Kerr black hole or some
other exotic object within GR. In particular, we consider a subclass of MN 
spacetimes characterized by three parameters (mass $M$,
spin angular momentum $J$ and mass quadrupole moment $Q$). In
a Kerr spacetime, the quadrupole moment $Q_{\rm Kerr}$ 
and all the higher order multipole moments are known to be
functions of the mass and spin. Thus, if $Q=Q_{\rm Kerr}(M,J)$
our MN spacetimes exactly reduce to a Kerr BH. This
property is very convenient because it makes these spacetimes
an ideal tool to set-up a null experiment to test the BH paradigm: any experiment pointing 
at a value of $Q$ significantly different from $Q_{\rm Kerr}(M,J)$ 
would imply that the object under consideration is 
either a compact object different from a BH within GR, or a BH or a compact object in
a gravity theory different from GR. In this sense, MN spacetimes can be used not
only to test the BH hypothesis but also to test the gravity theory itself.

The accretion process in our MN spacetimes can be qualitatively different 
than in a Kerr spacetime. In particular, we find that when the accreting gas
reaches the ISCO (i.e. the inner edge of the Novikov-Thorne disk model) 
there are four qualitatively different possibilities:
\begin{enumerate}
\item The ISCO is {\it radially} unstable, and the gas plunges into the  
compact object remaining roughly on the equatorial plane. This is the same scenario as in the Kerr case.
\item The ISCO is {\it radially} unstable and the gas 
plunges, but does not reach the compact object. Instead, 
it gets trapped between the object and the ISCO and 
forms a thick disk.
\item The ISCO is {\it vertically} unstable, and the gas plunges into the                                                                
compact object {\it outside} the equatorial plane.
\item The ISCO is {\it vertically} unstable and the gas 
plunges, but does not reach the compact object. Instead,
it gets trapped between the object and the ISCO and        
forms two thick disks, above and below the equatorial plane.
\end{enumerate}

This paper is organized as follows. In Sec.~\ref{s-mn} we review 
a subclass of the MN solutions, characterized by three free parameters $(M,J,Q)$,
that we use to describe the spacetime around generic 
compact objects. In Sec.~\ref{s-p}, we study plunging orbits in these 
MN spacetimes and we show that the accreting gas may not reach the 
surface of the compact object. This may lead to the formation of 
a thick disk inside the ISCO.
In Sec.~\ref{s-d} we discuss the possible astrophysical consequences of these thick disks. 
Finally, in Sec.~\ref{s-c} we draw our conclusions, while in
 Appendix~\ref{app} we review the theory of thick non-gravitating disks in axisymmetric stationary spacetimes, which we use to describe the thick disks inside the ISCO. In Appendix~\ref{app2} we review the thermal bremsstrahlung emission rate.

Throughout the paper we use units in which $G=c=1$, 
unless stated otherwise.

\section{Manko-Novikov spacetimes \label{s-mn}}

The MN metric~\cite{mn} is a stationary, axisymmetric, and asymptotically flat 
exact solution of the Einstein vacuum equations with arbitrary mass-multipole moments. In quasi-cylindrical coordinates $(\rho,z)$
and in prolate spheroidal coordinates $(x,y)$, the line element is, respectively,
\begin{widetext}
\be\label{eq-ds2}
ds^2 &=& - f \left(dt - \omega d\phi\right)^2
+ \frac{e^{2\gamma}}{f} \left(d\rho^2 + dz^2\right)
+ \frac{\rho^2}{f} d\phi^2 = \nonumber\\
&=& - f \left(dt - \omega d\phi\right)^2
+ \frac{k^2 e^{2\gamma}}{f}\left(x^2 - y^2\right)
\left(\frac{dx^2}{x^2 - 1} + \frac{dy^2}{1 - y^2}\right)
+ \frac{k^2}{f} \left(x^2 - 1\right)\left(1 - y^2\right) 
d\phi^2 \, ,
\ee
where
\be
f = e^{2\psi} A/B\, , \quad
\omega = 2 k e^{- 2\psi} C A^{-1} 
- 4 k \alpha \left(1 - \alpha^2\right)^{-1} \, , \quad
e^{2\gamma} &=& e^{2\gamma'}A \left(x^2 - 1\right)^{-1}
\left(1 - \alpha^2\right)^{-2} \, ,
\ee
and
\be
\psi &=& \sum_{n = 1}^{+\infty} \frac{\alpha_n P_n}{R^{n+1}} 
\, , \\\label{gammapdef}
\gamma' &=& \frac{1}{2} \ln\frac{x^2 - 1}{x^2 - y^2} 
+ \sum_{m,n = 1}^{+\infty} \frac{(m+1)(n+1) 
\alpha_m \alpha_n}{(m+n+2) R^{m+n+2}}
\left(P_{m+1} P_{n+1} - P_m P_n\right) + \nonumber\\
&& + \left[ \sum_{n=1}^{+\infty} \alpha_n 
\left((-1)^{n+1} - 1 + \sum_{k = 0}^{n}
\frac{x-y+(-1)^{n-k}(x+y)}{R^{k+1}}P_k \right) \right] \, , \\
A &=& (x^2 - 1)(1 + \tilde{a}\tilde{b})^2 - (1 - y^2)(\tilde{b} - \tilde{a})^2 \, , \\
B &=& [x + 1 + (x - 1)\tilde{a}\tilde{b}]^2 + [(1 + y)\tilde{a} + (1 - y)\tilde{b}]^2 \, , \\
C &=& (x^2 - 1)(1 + \tilde{a}\tilde{b})[\tilde{b} - \tilde{a} - y(\tilde{a} + \tilde{b})] 
+ (1 - y^2)(\tilde{b} - \tilde{a})[1 + \tilde{a}\tilde{b} + x(1 - \tilde{a}\tilde{b})] \, , \\\label{adef}
\tilde{a} &=& -\alpha \exp \left[\sum_{n=1}^{+\infty} 2\alpha_n 
\left(1 - \sum_{k = 0}^{n} \frac{(x - y)}{R^{k+1}} 
P_k\right)\right] \, , \\\label{bdef}
\tilde{b} &=& \alpha \exp \left[\sum_{n=1}^{+\infty} 2\alpha_n 
\left((-1)^n + \sum_{k = 0}^{n} \frac{(-1)^{n-k+1}(x + y)}{R^{k+1}} 
P_k\right)\right] \, .
\ee
\end{widetext}
Here $R = \sqrt{x^2 + y^2 - 1}$ and $P_n$ are the Legendre 
polynomials with argument $xy/R$,
\be
P_n &=& P_n\left(\frac{xy}{R}\right) \, , \nonumber\\
P_n(\chi) &=& \frac{1}{2^n n!} \frac{d^n}{d\chi^n} 
\left(\chi^2 - 1\right)^n \, ,
\ee
while the relation
between prolate spheroidal and quasi-cylindrical
coordinates is given by
\be
\rho = k \sqrt{\left(x^2 - 1\right)\left(1 - y^2\right)} \, ,
\qquad
z = kxy \, ,
\ee
with inverse
\begin{align}
&x = \frac{1}{2k} \left(\sqrt{\rho^2 + \left(z + k\right)^2}
+ \sqrt{\rho^2 + \left(z - k\right)^2}\right) \, , \nonumber\\
&y = \frac{1}{2k} \left(\sqrt{\rho^2 + \left(z + k\right)^2}
- \sqrt{\rho^2 + \left(z - k\right)^2}\right) \, .
\end{align}

The MN solution has an infinite number of free parameters: $k$, which regulates the 
mass of the spacetime; $\alpha$, which regulates the spin; and $\alpha_n$ ($n=1, . . . , +\infty$)
which regulates the mass-multipole moments, starting from the dipole $\alpha_1$, to the quadrupole  $\alpha_2$, etc. 
For $\alpha \neq 0$ 
and $\alpha_n = 0$, the MN solution reduces to the Kerr metric. For $\alpha = 
\alpha_n = 0$, it reduces to the Schwarzschild solution. For $\alpha = 0$ 
and $\alpha_n \neq 0$, one obtains the static Weyl metric.

The no-hair theorem~\cite{nohair} states that the only asymptotically flat, vacuum 
and stationary solution of the Einstein equations that is non-singular on and outside an event horizon
and that presents no closed timelike curves outside it is given by the Kerr metric. Therefore,
the MN spacetime must either have no event horizon, or present naked singularities or closed timelike curves outside it.
In fact, the surface $x = 1$ ($\rho=0$, $|z|\leq k$), which is the event horizon in the Kerr case $\alpha_n = 0$,
in general is only a partial horizon, because it presents a naked curvature singularity
on the equatorial plane (i.e. at $x=1$, $y=0$, corresponding to $\rho=z=0$)~\cite{mn}. Also, there 
are closed timelike curves outside it.
However, these pathological features appear at small radii and here 
the basic idea is that naked singularities and closed timelike 
curves do not exist in reality because they are either inside some sort of exotic compact
object, whose \textit{exterior} gravitational field is described by 
the MN metric, or because GR breaks down close to them; 
see e.g. Ref.~\cite{string} for some specific 
mechanisms that can do the job.

Without loss of generality, we can put $\alpha_1 = 0$ to bring 
the massive object to the origin of the coordinate system. In what 
follows, we restrict our attention to the subclass of MN spacetimes 
with $\alpha_n = 0$ for $n \neq 2$. We then have three free parameters 
($k$, $\alpha$, and $\alpha_2$) related to the mass $M$, the 
dimensionless spin parameter $a = J/M^2$, and the dimensionless 
anomalous quadrupole moment $q = -(Q - Q_{\rm Kerr})/M^3$, 
by the relations
\be
\alpha &=& \frac{\sqrt{1 - a^2} - 1}{a} \, , \\
k &=& M \frac{1 - \alpha^2}{1 + \alpha^2} \, , \\
\alpha_2 &=& q \frac{M^3}{k^3} \, .
\ee
Note that $q$ measures the deviation from the quadrupole
moment of a Kerr BH. In particular, since $Q_{\rm Kerr} = - a^2 M^3$, 
the solution is oblate for $q > - a^2$ and prolate for $q < - a^2$.
When $q=0$, the solution reduces to the Kerr metric, but when $q \neq 0$ also the higher-order mass-multipole 
moments have a different value than in Kerr.

Because the MN metric is stationary and axisymmetric, geodesic orbits have two constants of motion, the specific
energy  $E = -u_t$ and the 
$z$-component of the specific angular momentum
$L = u_\phi$. The $t$- and $\phi$-components of the 4-velocity 
of a test-particle are therefore 
\be
u^t &=& \frac{E g_{\phi\phi} + L g_{t\phi}}{
g_{t\phi}^2 - g_{tt} g_{\phi\phi}} \, , \\ 
u^\phi &=& - \frac{E g_{t\phi} + L g_{tt}}{
g_{t\phi}^2 - g_{tt} g_{\phi\phi}} \, .
\ee
From the normalization of the 4-velocity, $g_{\mu\nu}u^\mu 
u^\nu = -1$, we can write
\be
\frac{e^{2\gamma}}{f} \left[(u^\rho)^2+(u^z)^2\right] = V_{\rm eff}(E,L,\rho,z) \, ,
\ee
where the effective potential $V_{\rm eff}$ is defined by
\be
V_{\rm eff} = \frac{E^2}{f} - \frac{f}{\rho^2}
\left(L - \omega E\right)^2 - 1 \, .
\ee
Circular orbits in the equatorial plane must have  $\dot{\rho} = \dot{z} = 0$, 
which implies $V_{\rm eff} = 0$, and $\ddot{\rho} = \ddot{z} = 0$, which implies
$\partial_\rho V_{\rm eff} = 0$ and $\partial_z V_{\rm eff} = 0$. This means that circular equatorial
orbits are located at simultaneous zeros and extrema of the effective potential, where 
$\partial_\rho V_{\rm eff} =\partial_z V_{\rm eff} =  V_{\rm eff}=0$.
Because  $\partial_z V_{\rm eff} = 0$ is satisfied identically for $z=0$ (simply because of the reflection
symmetry of the MN metric with respect to the equatorial plane), from these conditions one can obtain $E$ and 
$L$ as a function of the radius $r$ of the circular equatorial orbit. These orbits are stable under small perturbations
in the radial direction if 
$\partial_\rho^2 V_{\rm eff} < 0$, and in the vertical direction if $\partial_z^2 V_{\rm eff} < 0$.

\begin{figure}
\par
\begin{center}
\includegraphics[type=pdf,ext=.pdf,read=.pdf,width=8.8cm]{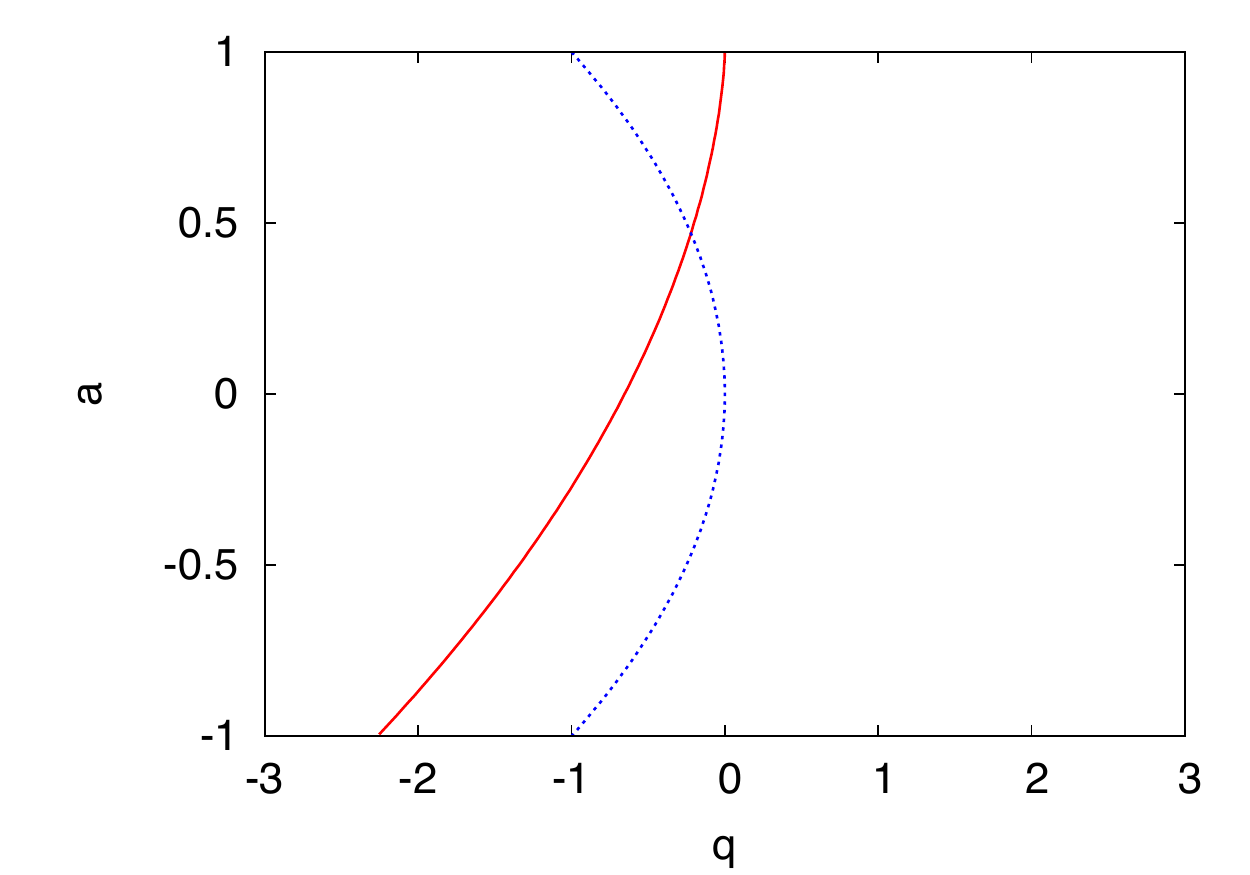}
\end{center}
\par
\vspace{-5mm} 
\caption{The character of the effective ISCO for MN spacetimes with spin parameter $a$
and anomalous quadrupole moment $q$. Compact objects to the 
right (left) of the solid red line have ISCOs determined by 
the instability in the radial (vertical) direction. Compact
objects to the right (left) of the dotted blue line are oblate
(prolate) bodies, while they are more oblate (prolate) than
a BH if $q > 0$ ($q < 0$).}
\label{f-2-1}
\end{figure}

As far as the position of the ISCO is concerned, for any given value of $a$
there are two critical values $q_1$ and $q_2$ for the anomalous quadrupole moment 
(both negative, $q_2<q_1<0$, but whose exact values depend on the spin $a$).
For $q \ge q_1$, circular orbits on the equatorial plane are always
vertically stable and the ISCO radius is determined by the onset of the orbital instability
in the radial direction (i.e. the ISCO is the marginally stable circular orbit in the radial
direction). This is the picture that one is familiar with in the Kerr case, which is indeed included in this range of
$q$'s (since for Kerr $q=0$).

For $q_2<q<q_1$, there are two circular orbits 
$r=r_1$ and $r=r_2$ (with $r_1>r_2$)
that are vertically stable but only marginally stable in the radial direction,
and one circular orbit $r=r_3$ (with $r_3<r_2$)  that is radially stable but only marginally stable in the vertical direction.
Stable circular orbits therefore exist for $r>r_1$ and for $r_3<r<r_2$, while orbits 
with $r_2<r<r_1$ are radially unstable (although vertically stable)
and orbits with $r<r_3$ are vertically unstable (although radially stable). 
As far as an accretion disk is concerned, however, what is relevant
is the radius $r=r_1$ of the outer marginally stable circular orbit, because that is
the radius at which the gas starts plunging.
We thus dub the circular orbit at $r=r_1$ the ``effective'' ISCO.
It is important to notice, however, that the stable orbits in
the ``inner'' region $r_3<r<r_2$ do \textit{not} necessarily have energy and angular momentum larger than those 
of the effective ISCO. For this reason, plunging orbits starting at the effective ISCO may hit a potential
barrier preventing them from reaching the compact object, at least in some regions of the parameter space $(a,q)$.
We will study this situation in detail in the next section.

As $q$ decreases towards $q_2$, the values of $r_1$ and 
$r_2$ approach and eventually coincide for $q=q_2$. For $q<q_2$,
there are no marginally stable orbits in the radial direction, i.e. all circular orbits
are radially stable. However, the marginally stable orbit in the vertical direction, $r=r_3$, still exists
and marks the position of the ISCO, which is therefore determined by the onset of the vertical instability.

In conclusion, for $q>q_2(a)$ [with $q_2(a)<0$], the effective ISCO for quasi-circular inspirals is given 
by the onset of the radial instability,
while for  $q<q_2(a)$ it is determined by the onset of the vertical instability. Also, let us note  that the critical value $q_2$ may be larger than
$-a^2$, that is, the ISCO can be determined by the vertical instability even 
for oblate objects (although since $q_2<0$ these objects need to be less oblate than a Kerr BH).
This situation is summarized in Fig.~\ref{f-2-1}, where we show the regions in the $(a,q)$ plane where compact
objects are prolate/oblate and more prolate/more oblate than a Kerr BH, and the regions where the effective ISCO is determined by
the vertical/radial instability.

In the rest of this paper we will present our results for MN spacetimes in terms
of the standard Boyer-Lindquist coordinates
$(r,\theta)$, which are related to the prolate
spheroidal coordinates $(x,y)$ and the quasi-cylindrical coordinates
$(\rho,z)$ used in this section by
\be
\rho &=& \sqrt{r^2 - 2 M r + a^2 M^2} \sin\theta \, , 
\nonumber\\
z &=& (r - M) \cos\theta \, .
\ee
and
\be
&& r = k x + M \, , 
\nonumber\\
&& \cos\theta = y \, .
\ee
Also, hereafter we will use coordinates in which the mass $M$ of the MN metric is 1.

\begin{figure*}
 \begin{tabular}{m{0.33\textwidth}m{0.33\textwidth}m{0.33\textwidth}} 
 		\includegraphics[type=pdf,ext=.pdf,read=.pdf,width=0.3\textwidth]{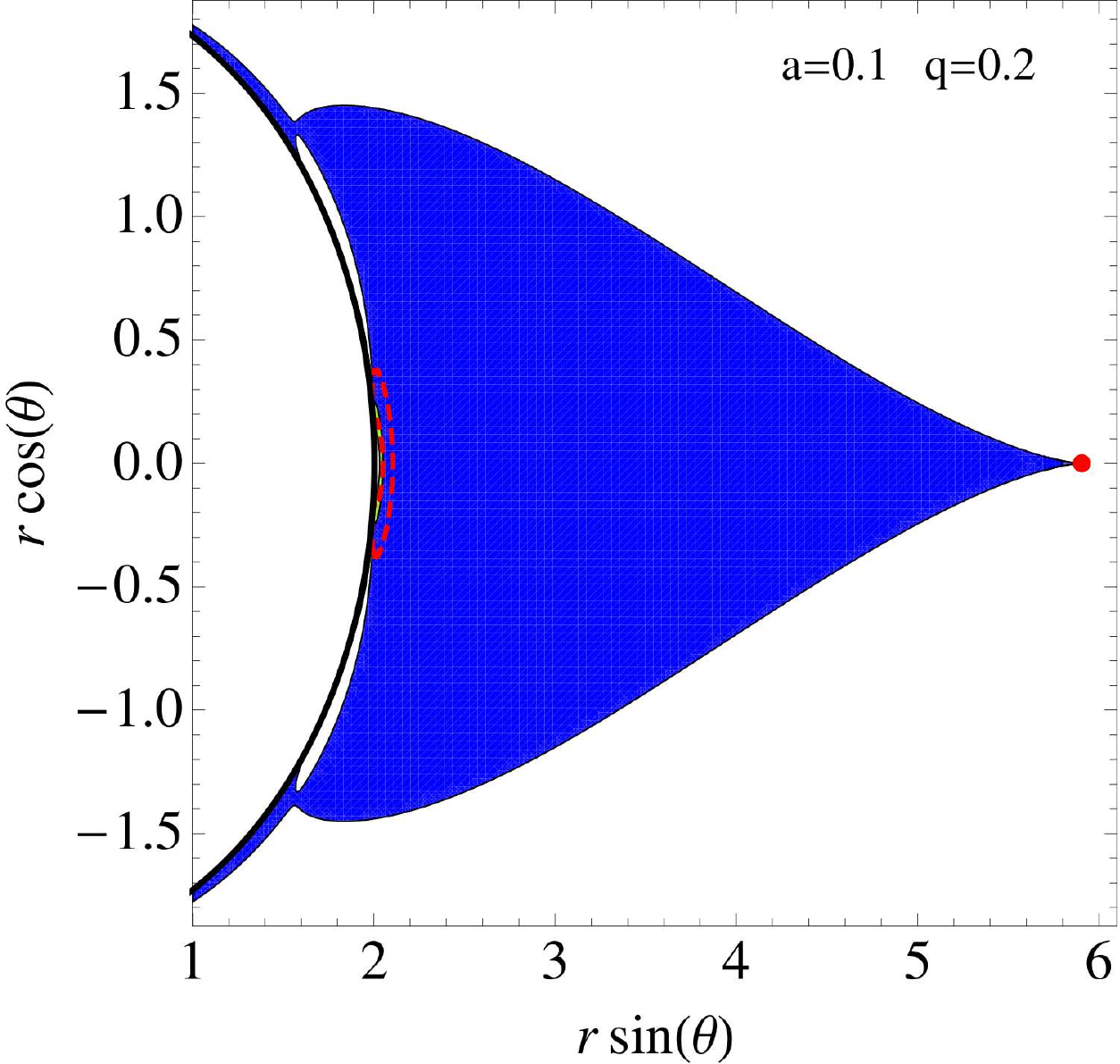} &
  		\includegraphics[type=pdf,ext=.pdf,read=.pdf,width=0.31\textwidth]{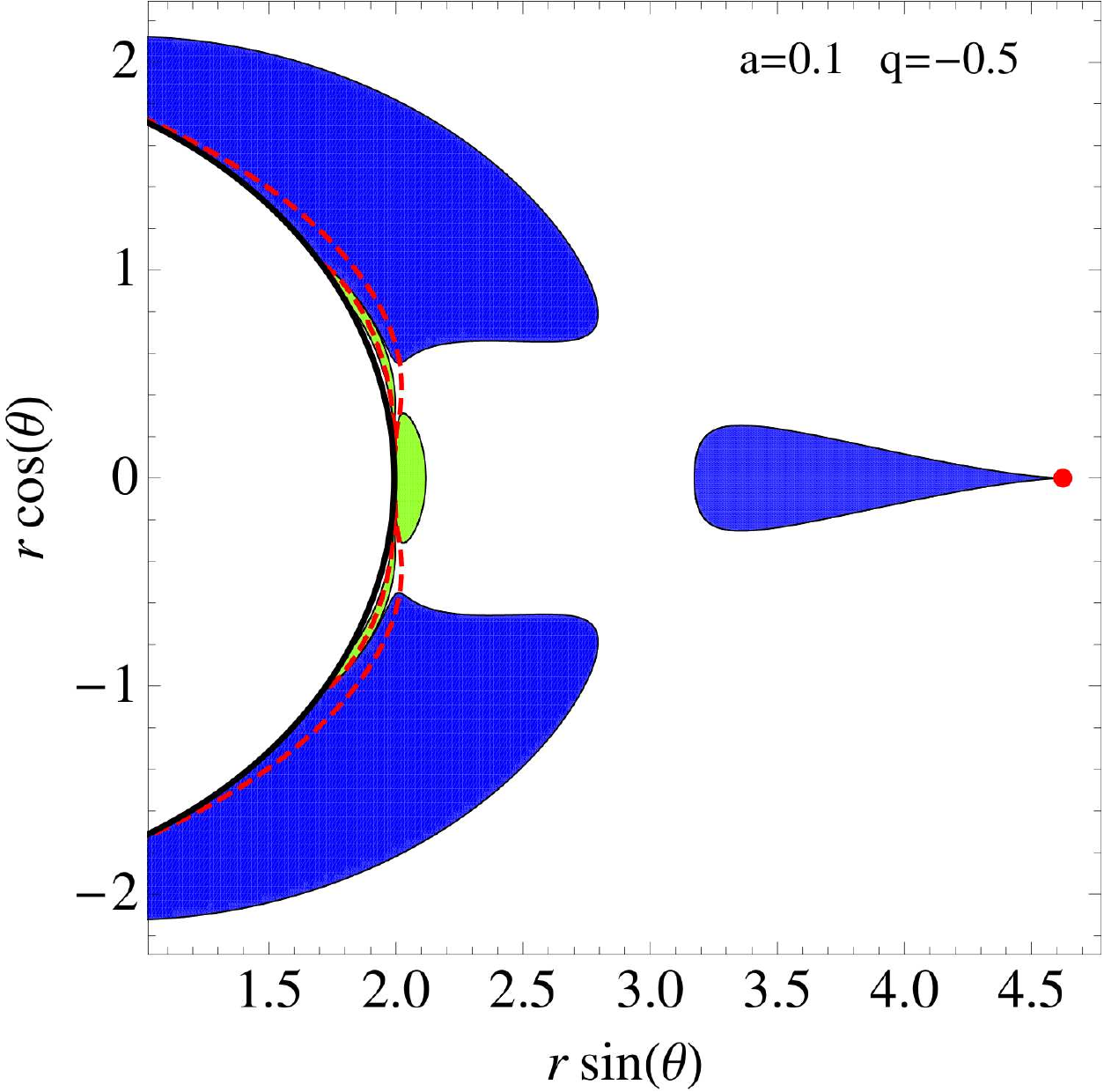} &
  		\includegraphics[type=pdf,ext=.pdf,read=.pdf,width=0.31\textwidth]{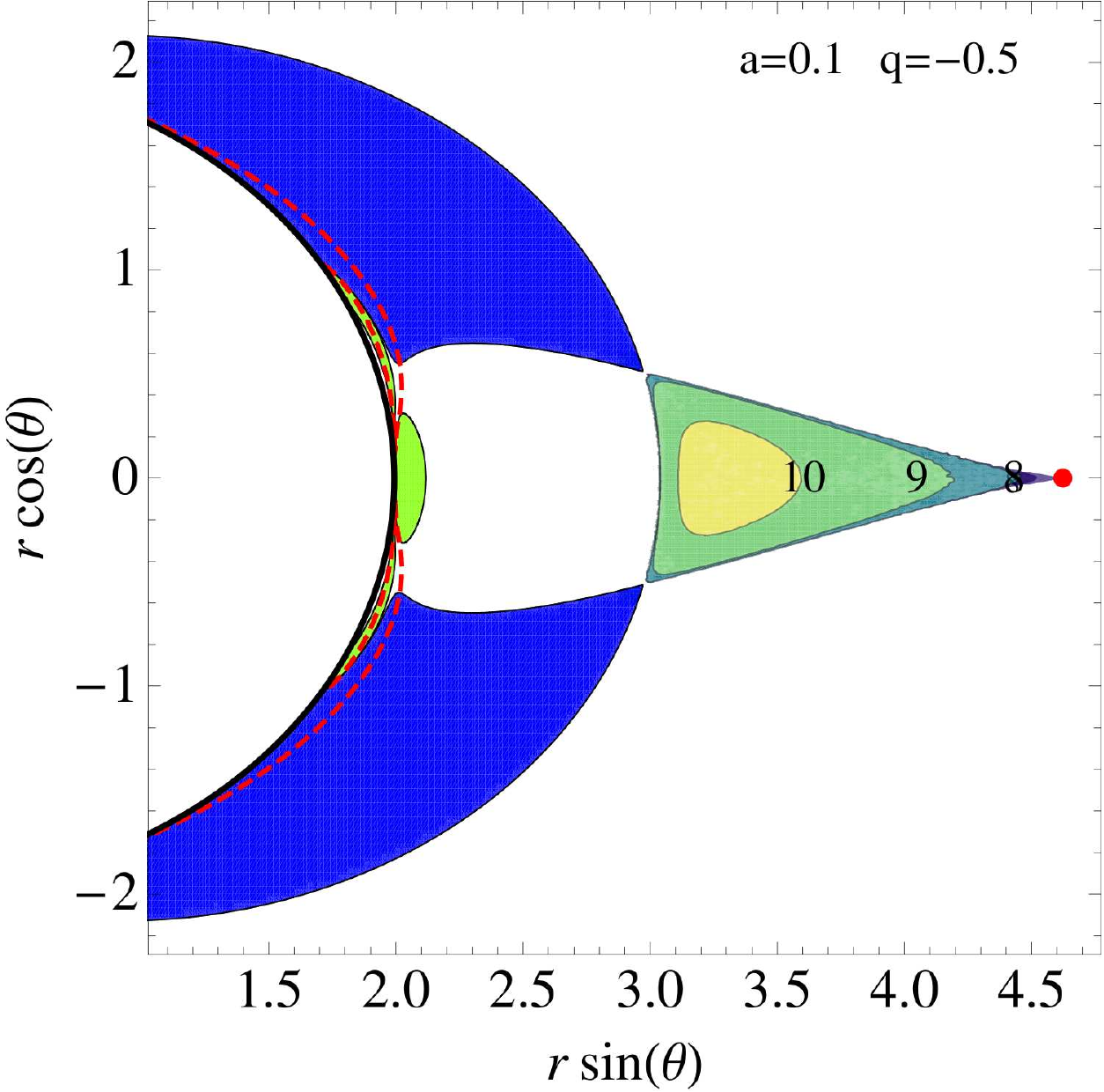} \\
\vskip1cm
 \begin{tabular}{m{0.16\textwidth}m{0.33\textwidth}m{0.31\textwidth}m{0.17\textwidth}}
											  &
  \includegraphics[type=pdf,ext=.pdf,read=.pdf,width=0.31\textwidth]{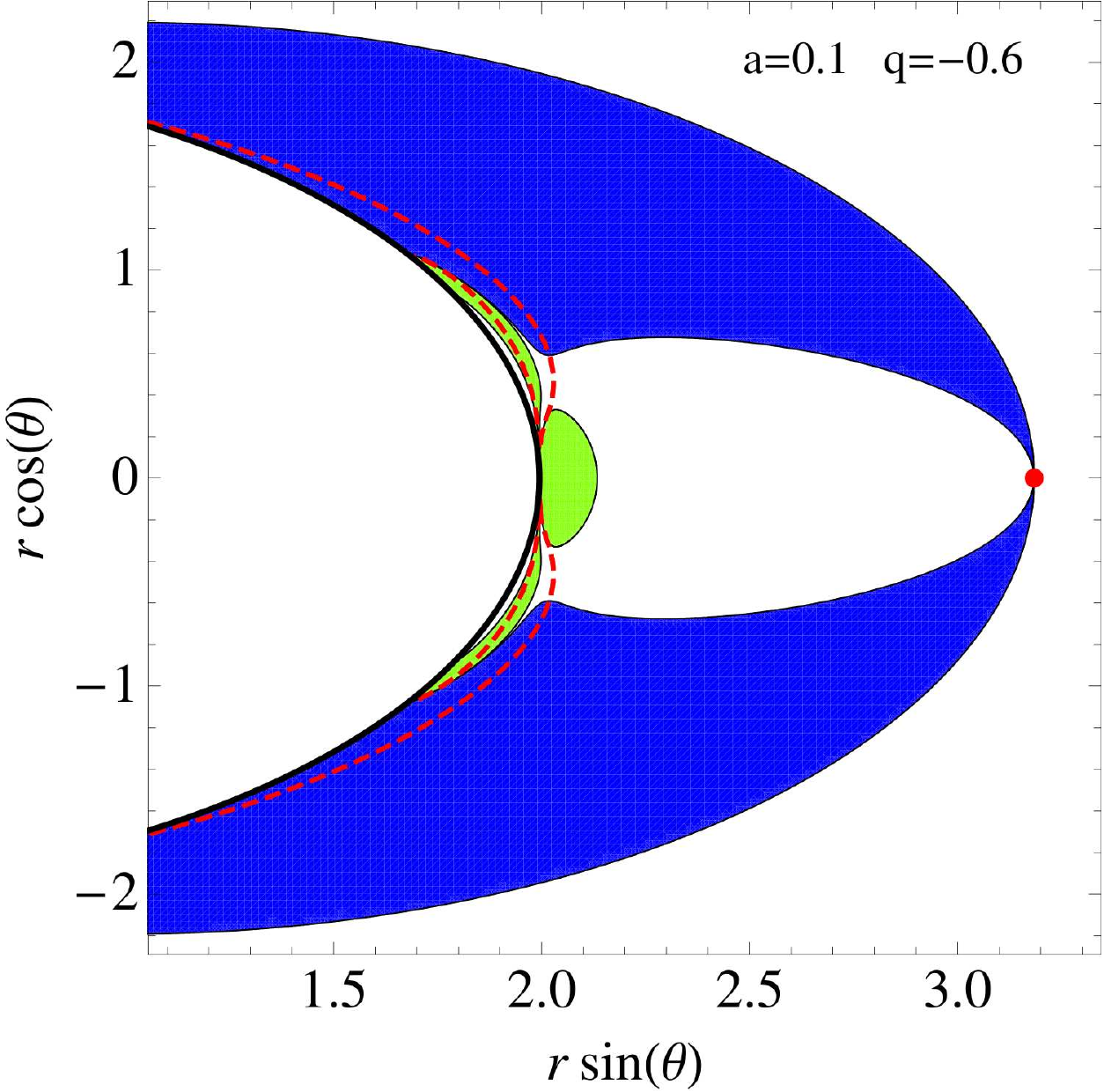} & 
  \includegraphics[type=pdf,ext=.pdf,read=.pdf,width=0.31\textwidth]{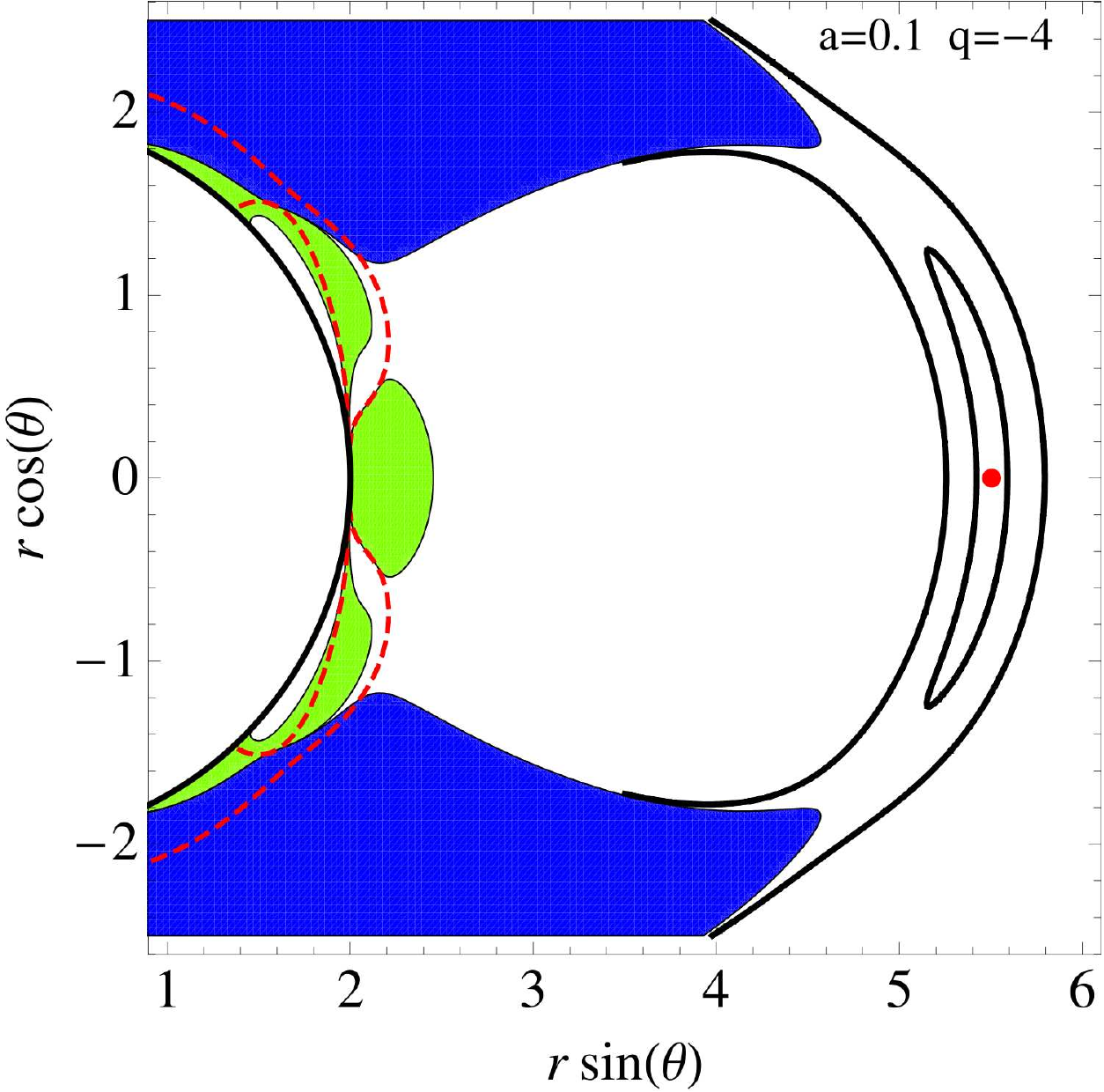} &\\
\\
 \end{tabular}
 \\
 \end{tabular}
\caption{Accretion in Manko-Novikov spacetimes with $a = 0.1$  and  $q=0.2$, $-0.5$, $-0.6$ and $q=-4$.
The solid black line at small radii
is the partial horizon $x=1$,  the region where closed timelike curves exist
(i.e. the region where $g_{\phi\phi} < 0$) is shown in green/light gray,
the red dashed line is the
boundary of the ergoregion (i.e. $g_{tt}=0$ on that line), while
the red dot on the equatorial plane marks the position of the ISCO, i.e. the inner edge of the thin accretion disk.
The thick disk, if it forms, sheds from the ISCO
and is denoted by concentric contours, whose label is an upper limit to the $\log_{\rm 10}$ of the gas temperature in K (see text for details).
In  blue/dark grey is  
the  region accessible to the gas shedding from the inner edges
of the thick disk [i.e. the region where $V_{\rm eff}(E_{\rm inner},L_{\rm inner},r,\theta)\geq0$, 
$E_{\rm inner}$ and $L_{\rm inner}$ being the energy and angular momentum of the gas at the inner edges of the thick disk].
If no thick disk
is present, in  blue/dark grey is the region accessible to the gas shedding from the inner edge of the thin disk 
 [i.e. the region where $V_{\rm eff}(E_{\rm ISCO},L_{\rm ISCO},r,\theta)\geq0$]: if this ``plunge'' region does not reach the ISCO,
we also show (with solid black lines around the ISCO) the 
contours $V_{\rm eff}(E_{\rm ISCO},L_{\rm ISCO},r,\theta)=-10^{-4}$ and  $V_{\rm eff}(E_{\rm ISCO},L_{\rm ISCO},r,\theta)=-10^{-3}$,
which  represent the region where the gas reaching the ISCO can shed if subject to a small perturbation.
\\
 For $q=0.2$ the ISCO is radially unstable, and the gas plunges directly into the  
 compact object remaining roughly on the equatorial plane, as in the Kerr case [scenario (1a)].
 For $q=-0.5$ the ISCO is radially unstable and the gas 
 plunges, but does not reach the compact object; instead, 
 it gets trapped between the object and the ISCO and 
 forms a thick disk [scenario (1b)].
 For $q=-0.6$ the ISCO is vertically unstable, and the gas plunges directly into the                                                                
 compact object outside the equatorial plane [scenario (2a)].
 For $q=-4$ the ISCO is vertically unstable, and the gas does not plunge directly but remains
trapped in the vicinities of the ISCO 
[because $V_{\rm eff}(E_{\rm ISCO},L_{\rm ISCO}, r,\theta)<0$ near the ISCO]. 
However, the potential barrier is not tall enough to allow a thick disk to form,
and a small perturbation is enough to cause the gas to fall into the central object [scenario (2b)].\label{fig_a_0.1}
}
\end{figure*}

\section{The plunge and the formation of thick disks inside the ISCO\label{s-p}}

At the inner edge of the thin accretion disk, which corresponds to the effective ISCO defined in the previous section,
the gas is expected to shed and plunge towards the central object. 
To understand the geometry of this plunge, it is convenient to plot the region 
accessible to the shedding gas, i.e. the region
accessible to particles having the ISCO specific energy and angular momentum $E_{\rm ISCO}$ and $L_{\rm ISCO}$.
This ``plunge'' region is defined by
\be\label{eq-pl}
V_{\rm eff} (E_{\rm ISCO}, L_{\rm ISCO}, r, \theta) \ge 0 \,,
\ee
and automatically includes the ISCO ($r=r_{\rm ISCO}$, $\theta=\pi/2$), but not necessarily a neighborhood of it. More specifically,
one can have several scenarios:
\begin{enumerate}
\item If the effective ISCO is determined by the onset of the radial instability, there are two possibilities:
\begin{enumerate}
\item The gas plunges directly into the compact object. As in the Kerr case, the ``plunge'' region and therefore the 
gas may expand above and below the equatorial plane
forming a sort of ``plume'', but they remain confined near the equatorial plane.
\item The gas starts to plunge from the ISCO, but gets trapped before reaching the compact object. 
\end{enumerate}
\item If the effective ISCO is determined by the onset of the vertical instability, there are three possibilities:
\begin{enumerate}
\item The gas plunges directly into the compact object. However, because the ISCO is stable in the radial direction, 
the gas does not shed on the equatorial plane as in the Kerr case, but sheds above and below the equatorial plane, forming two separate
``plumes''.
\item The ``plunge'' region does not contain a neighborhood of the ISCO, because 
$V_{\rm eff} (E_{\rm ISCO}, L_{\rm ISCO}, r, \theta)$ is strictly negative around it. However, a small perturbation 
(e.g. a small initial velocity) is enough to allow the gas to escape
the potential well surrounding the ISCO, thus shedding above and below the 
equatorial plane and plunging into the compact object in two separate plumes. Note that this case is therefore very similar to scenario (2a).
\item The gas sheds above and below the equatorial plane from the ISCO, but gets trapped before reaching the compact object.
\end{enumerate}
\end{enumerate}

Scenarios (1b) and (2c), where the gas sheds from the inner edge of the thin disk 
but gets trapped before reaching the compact object, are clearly non-stationary configurations,
because the gas keeps accumulating between the object and the ISCO. For the system to settle onto
a stationary configuration, the gas would have to overflow the potential barrier separating it from the object,
but this can be achieved only if the gas forms, inside the ISCO, a coherent structure that is
capable of further dissipating energy  and angular momentum (e.g. through
viscosity or magnetic fields).

\begin{figure*}
 \begin{tabular}{m{0.33\textwidth}m{0.33\textwidth}m{0.33\textwidth}}
\includegraphics[type=pdf,ext=.pdf,read=.pdf,width=0.31\textwidth]{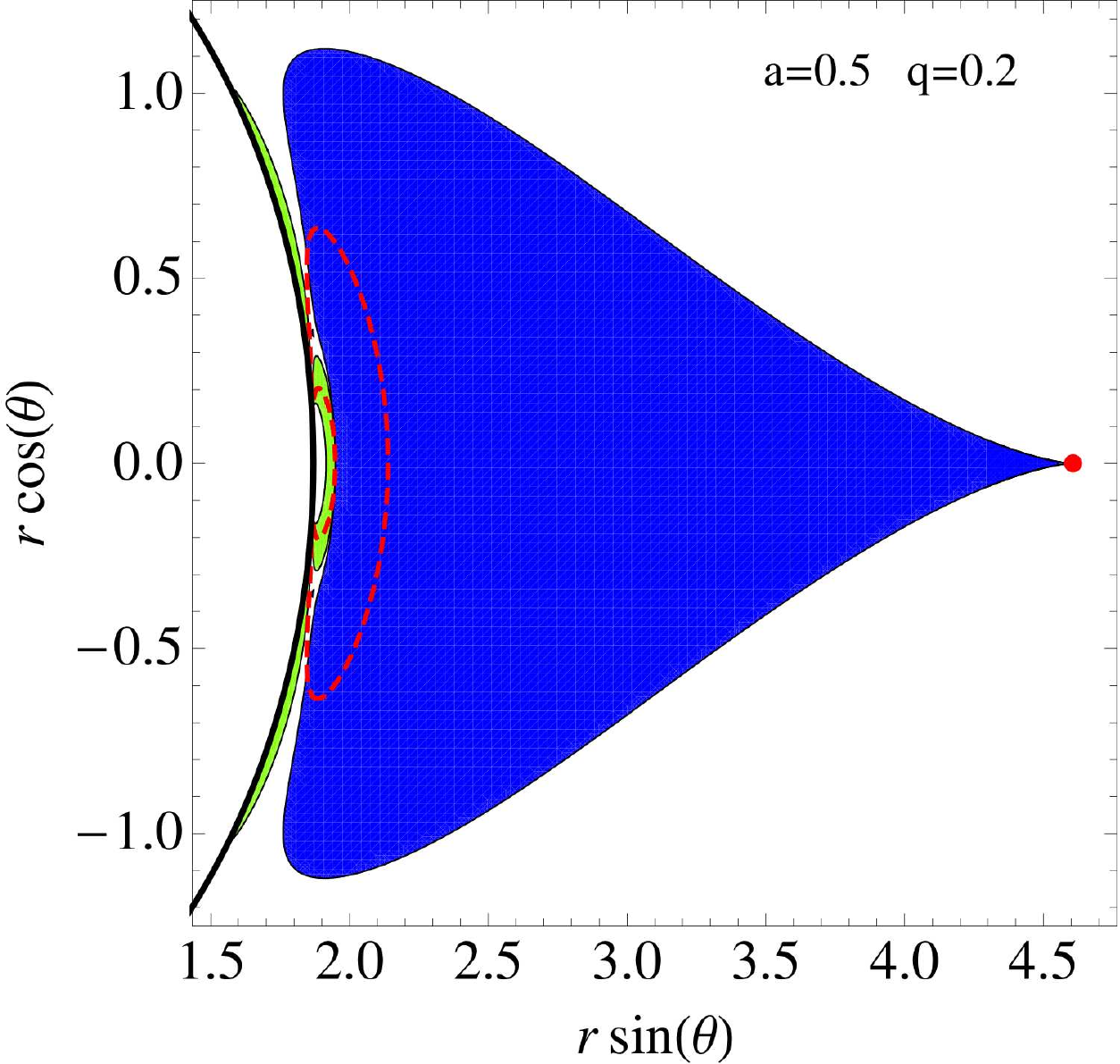}  &
                \includegraphics[type=pdf,ext=.pdf,read=.pdf,width=0.31\textwidth]{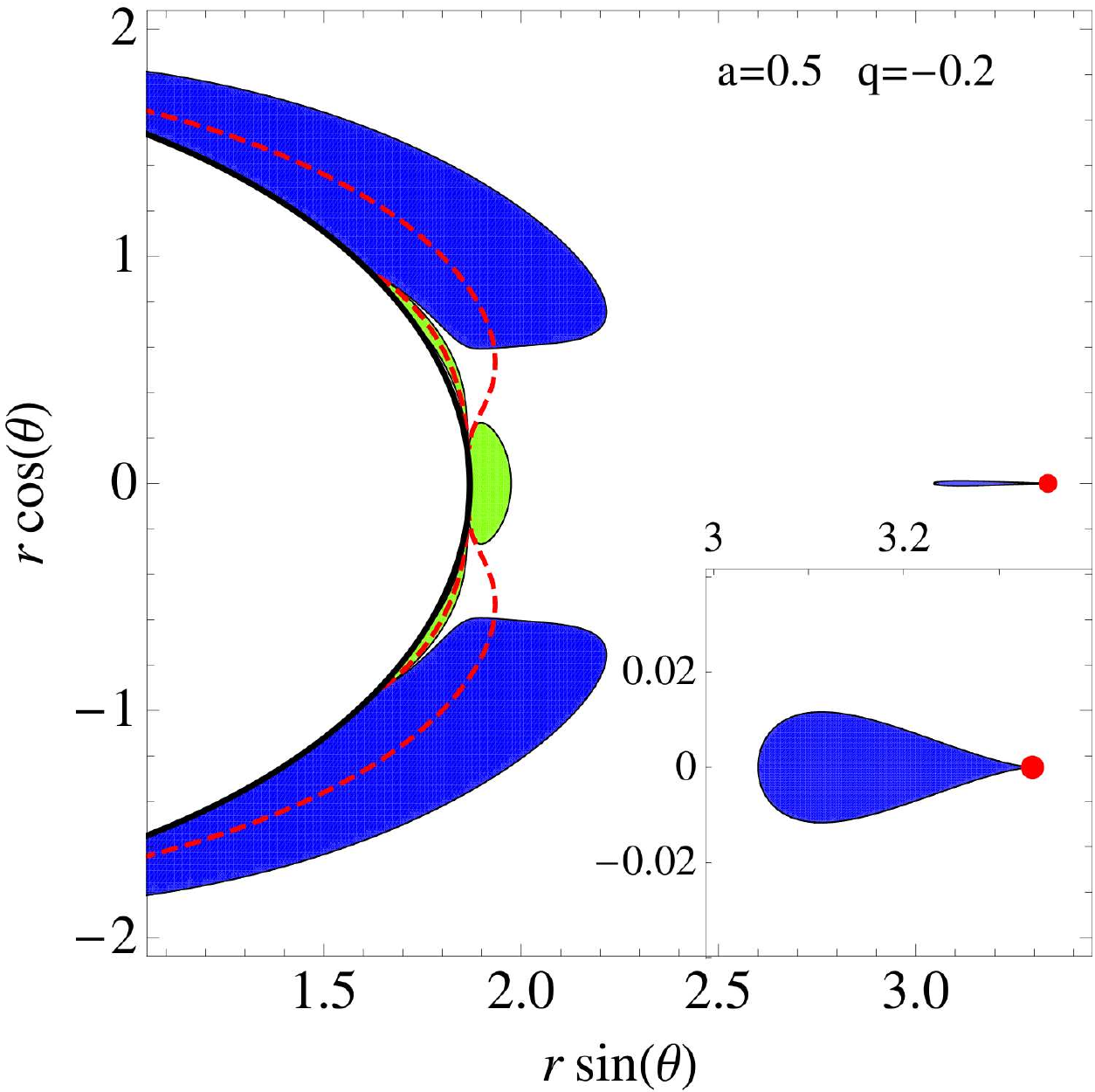} &
             \includegraphics[type=pdf,ext=.pdf,read=.pdf,width=0.31\textwidth]{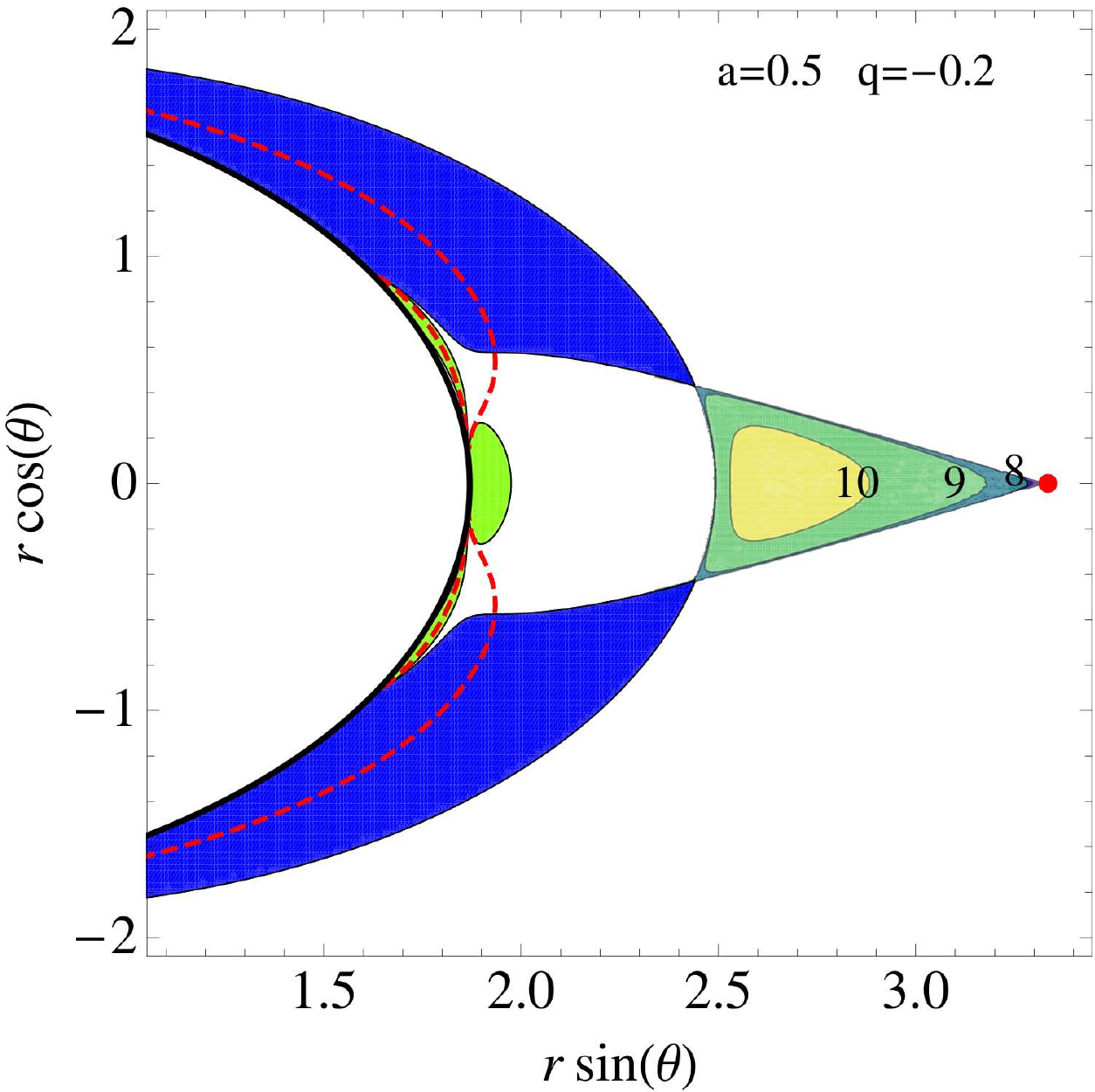} \\
\vskip1cm
 \begin{tabular}{m{0.16\textwidth}m{0.33\textwidth}m{0.33\textwidth}m{0.17\textwidth}} &  
  \includegraphics[type=pdf,ext=.pdf,read=.pdf,width=0.31\textwidth]{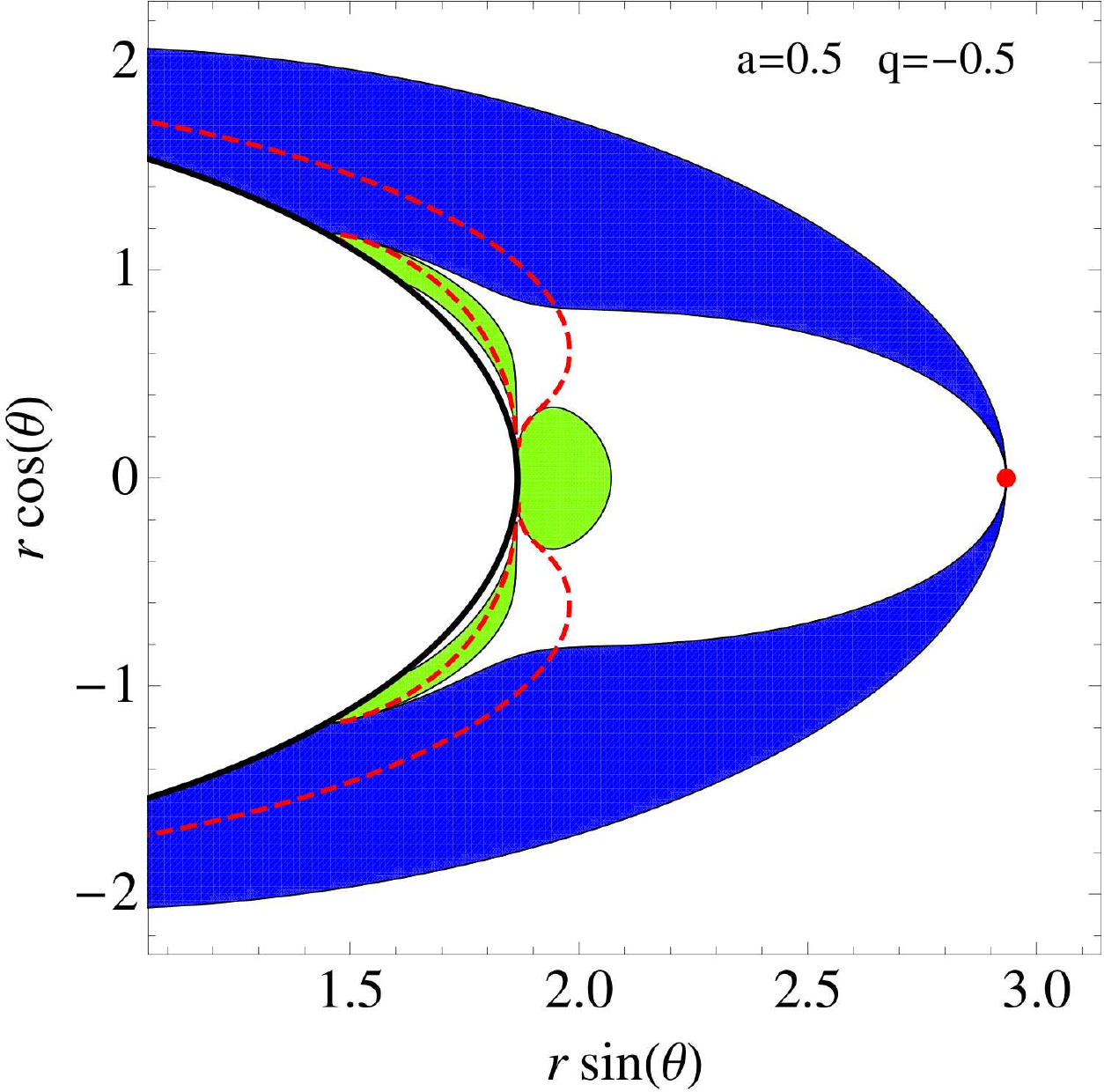} &
\includegraphics[type=pdf,ext=.pdf,read=.pdf,width=0.31\textwidth]{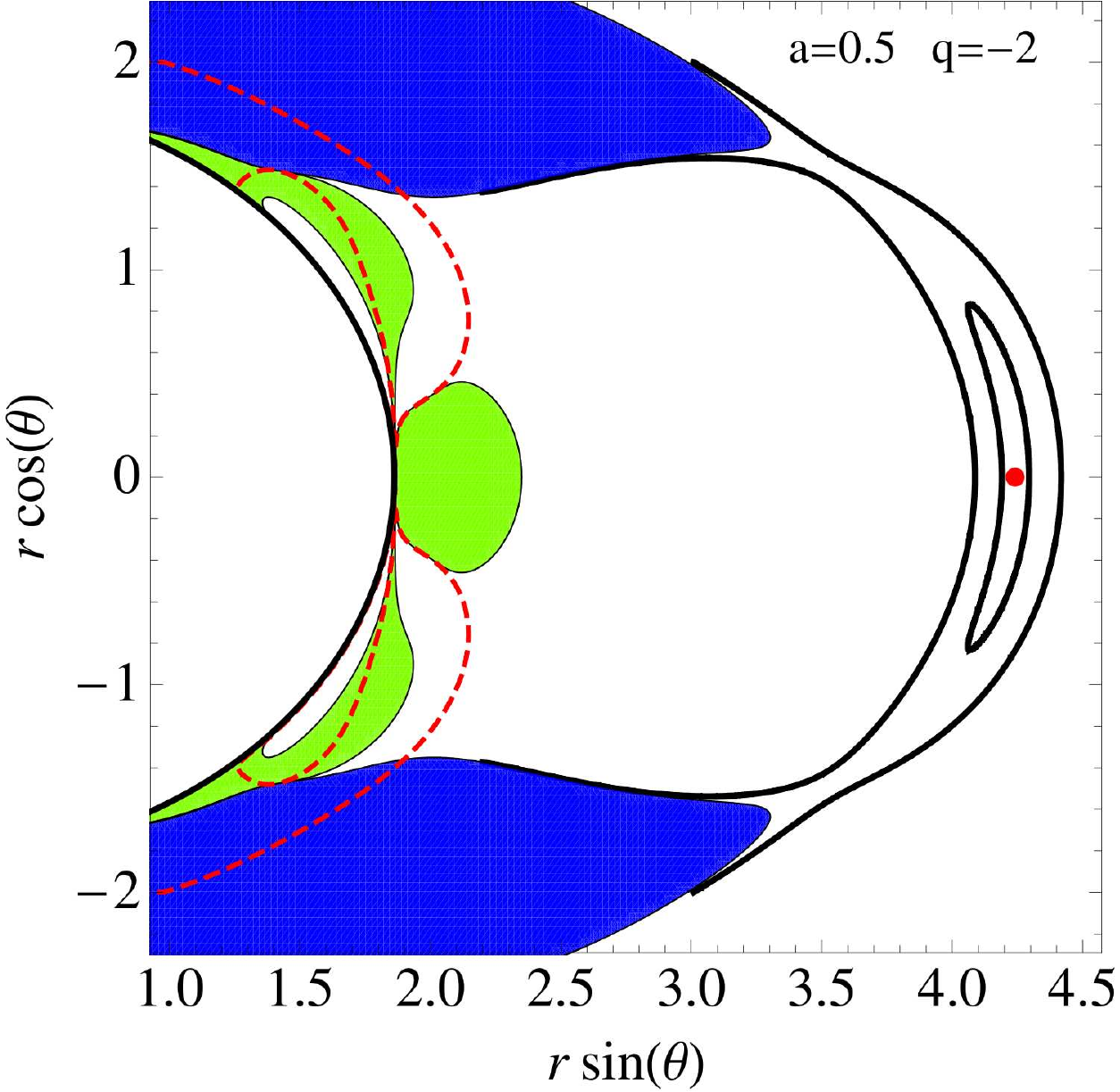} &\\
 \end{tabular}
 \\
 \end{tabular}
\caption{Same notation as in Fig.~\ref{fig_a_0.1}, but for $a = 0.5$ and $q=0.2$, $-0.2$, $-0.5$ and $-2$.
The accretion properties too are qualitatively similar, case by case, to those displayed for $a = 0.1$ in Fig.~\ref{fig_a_0.1}.
\label{fig_a_0.5}}
\end{figure*}

In this paper we study these ``coherent structures'' inside the ISCO using the theory
of non self-gravitating, stationary and axisymmetric thick disks developed in 
Refs.~\cite{pol1,pol2} (see also Refs.~\cite{hydrodrag,font_daigne,zanotti_etal}). 
We will briefly review this formalism in Appendix~\ref{app}, but for the present discussion it is sufficient
to mention that a thick disk is completely determined by specifying its inner or outer edge and the angular momentum distribution
on the equatorial plane. More precisely, one needs to assume a certain equatorial distribution for
the angular momentum per unit energy $\ell= L/E$, and this distribution completely determines
the distribution $\ell(r,\theta)$ of the angular momentum per unit energy outside the equatorial plane (see Appendix~\ref{app} for details).
Assuming in particular a power law, we can write
\be\label{eq-leq}
\bar{\ell}(r)\equiv\ell(r,\theta=\pi/2) = 
\ell_{\rm ISCO} \left(\frac{r}{r_{\rm ISCO}}\right)^\beta \,,
\ee
where $\ell_{\rm ISCO}=L_{\rm ISCO}/E_{\rm ISCO}$ and $\beta$ is a free parameter, which needs to be positive
in order for the thick disk to be stable~\cite{font_daigne}. 

In our case, we impose that the outer edge of the thick disk coincides with the ISCO (so that the thick disk is 
fed by the gas shedding from the thin disk), and we choose the parameter $\beta$ such that the thick disk also presents
an inner shedding point. This requirement completely fixes the value of $\beta$ to a certain critical value $\beta_{\rm crit}$.
The gas shedding from the inner edge of the thick disk will then plunge into the compact object, and the ``plunge'' region accessible 
to this gas will be described by 
\be
V_{\rm eff} (E_{\rm inner}, L_{\rm inner}, r, \theta) \ge 0 \, ,
\ee
where the specific energy and angular momentum of the gas at the inner edge of the thick disk,
$E_{\rm inner}$ and $L_{\rm inner}$, are determined by solving simultaneously $L_{\rm inner}/E_{\rm inner}=\ell(r_{\rm inner},\theta_{\rm inner})$ 
and $V_{\rm eff} (E_{\rm inner}, L_{\rm inner}, r_{\rm inner},\theta_{\rm inner})=0$.

\begin{figure*}
 \begin{tabular}{m{0.33\textwidth}m{0.33\textwidth}m{0.33\textwidth}} 
             \includegraphics[type=pdf,ext=.pdf,read=.pdf,width=0.32\textwidth]{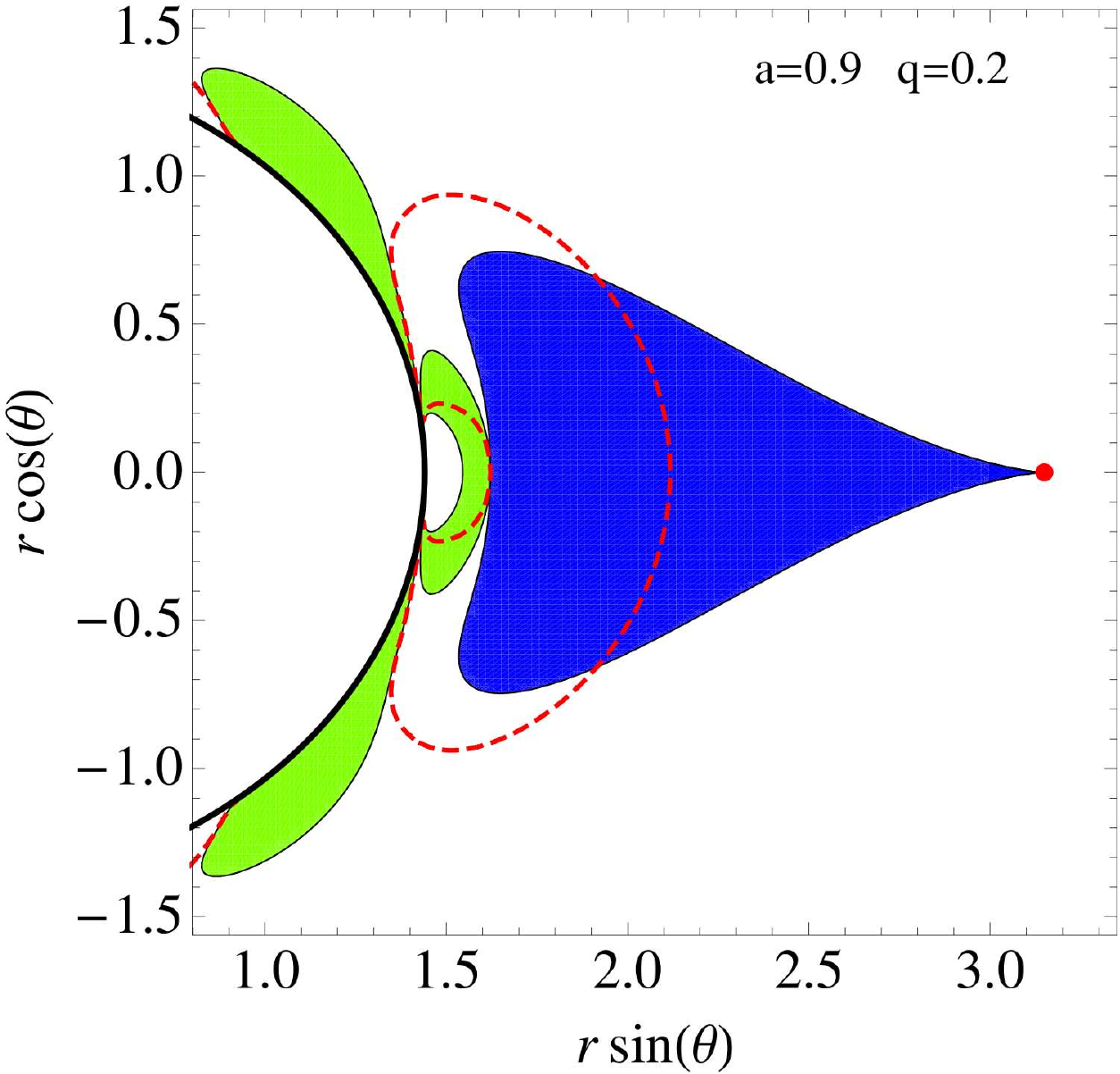}&
                \includegraphics[type=pdf,ext=.pdf,read=.pdf,width=0.32\textwidth]{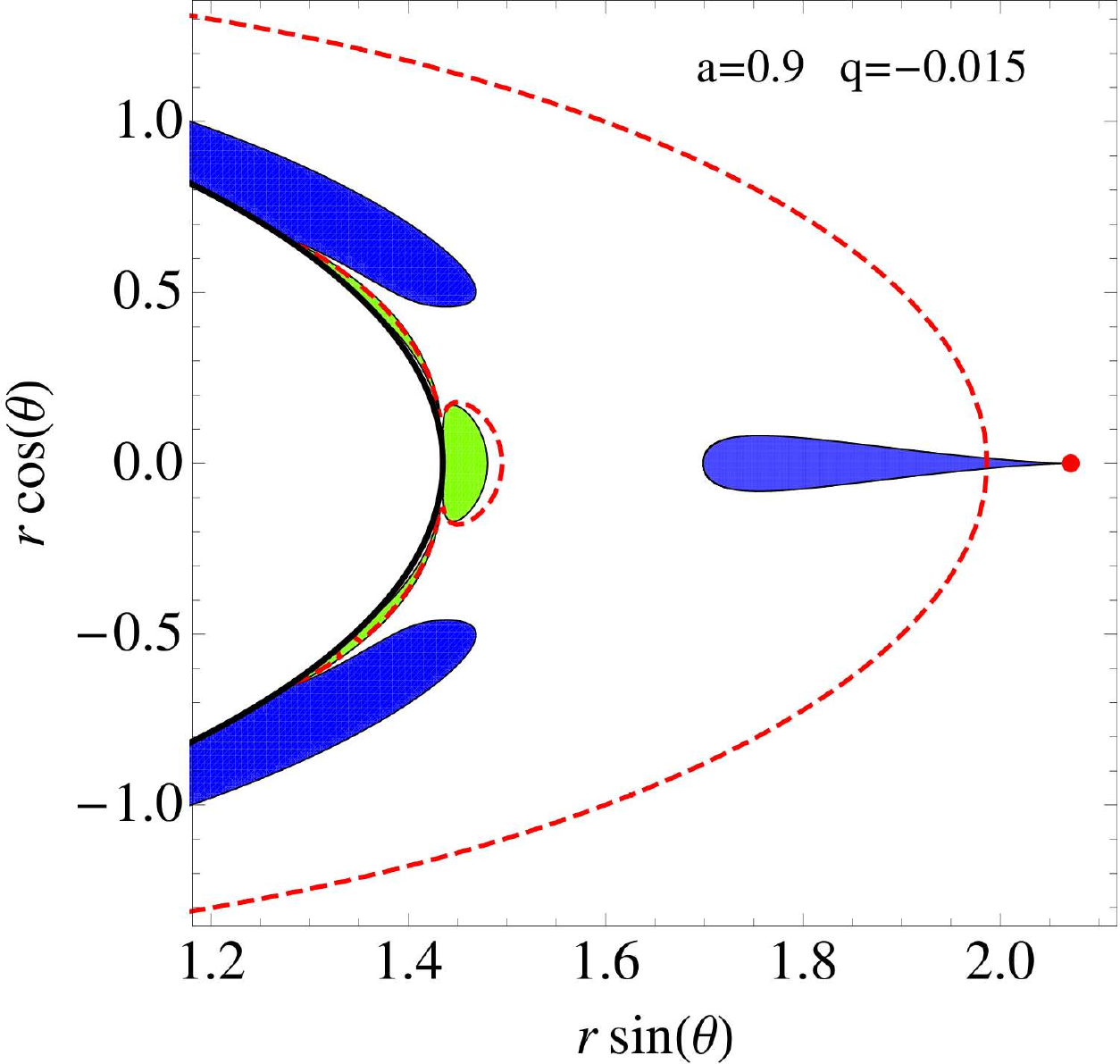} &
                \includegraphics[type=pdf,ext=.pdf,read=.pdf,width=0.32\textwidth]{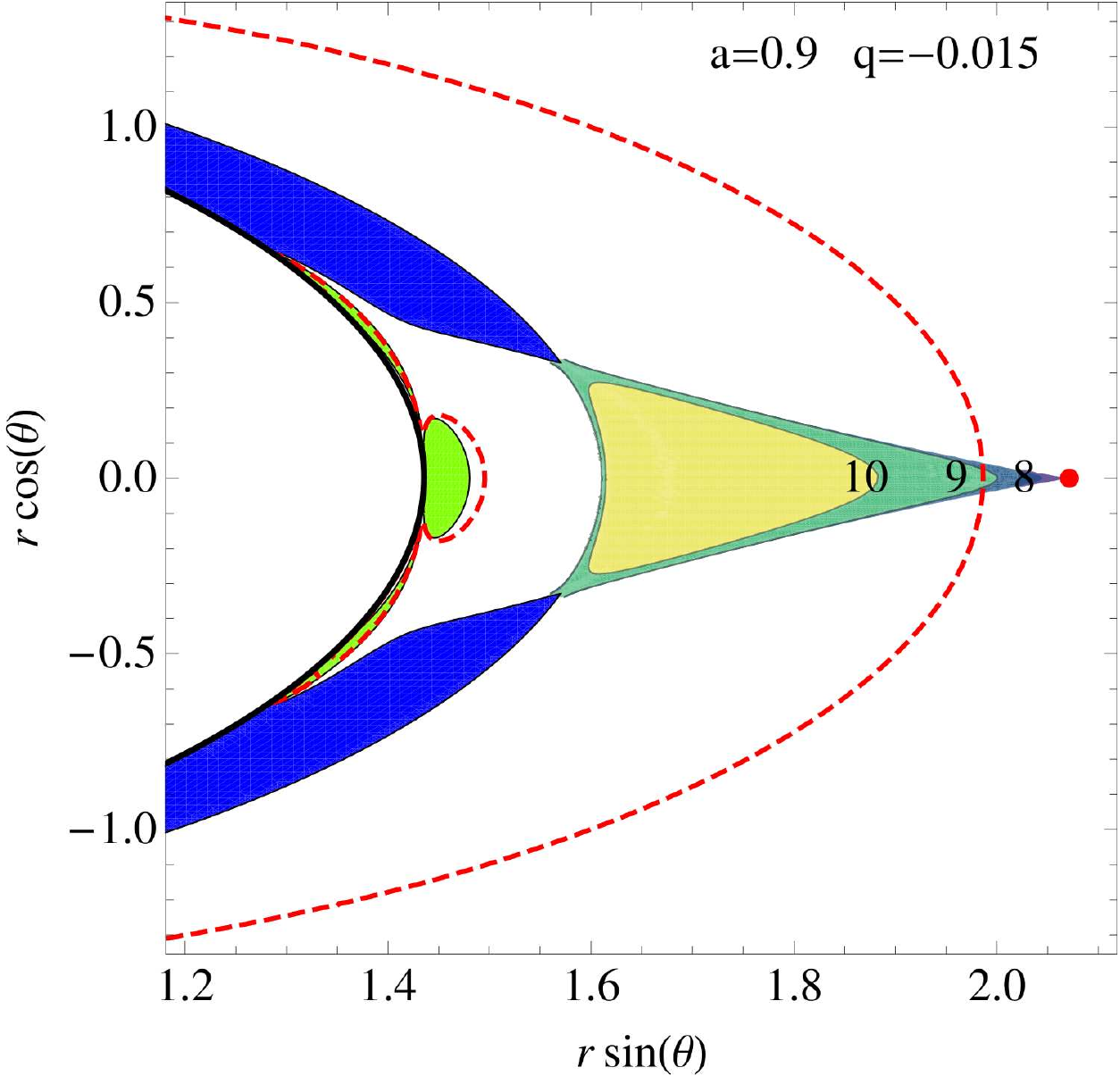} 
\\
\vskip1cm
 \begin{tabular}{m{0.33\textwidth}m{0.33\textwidth}m{0.33\textwidth}}
  \includegraphics[type=pdf,ext=.pdf,read=.pdf,width=0.32\textwidth]{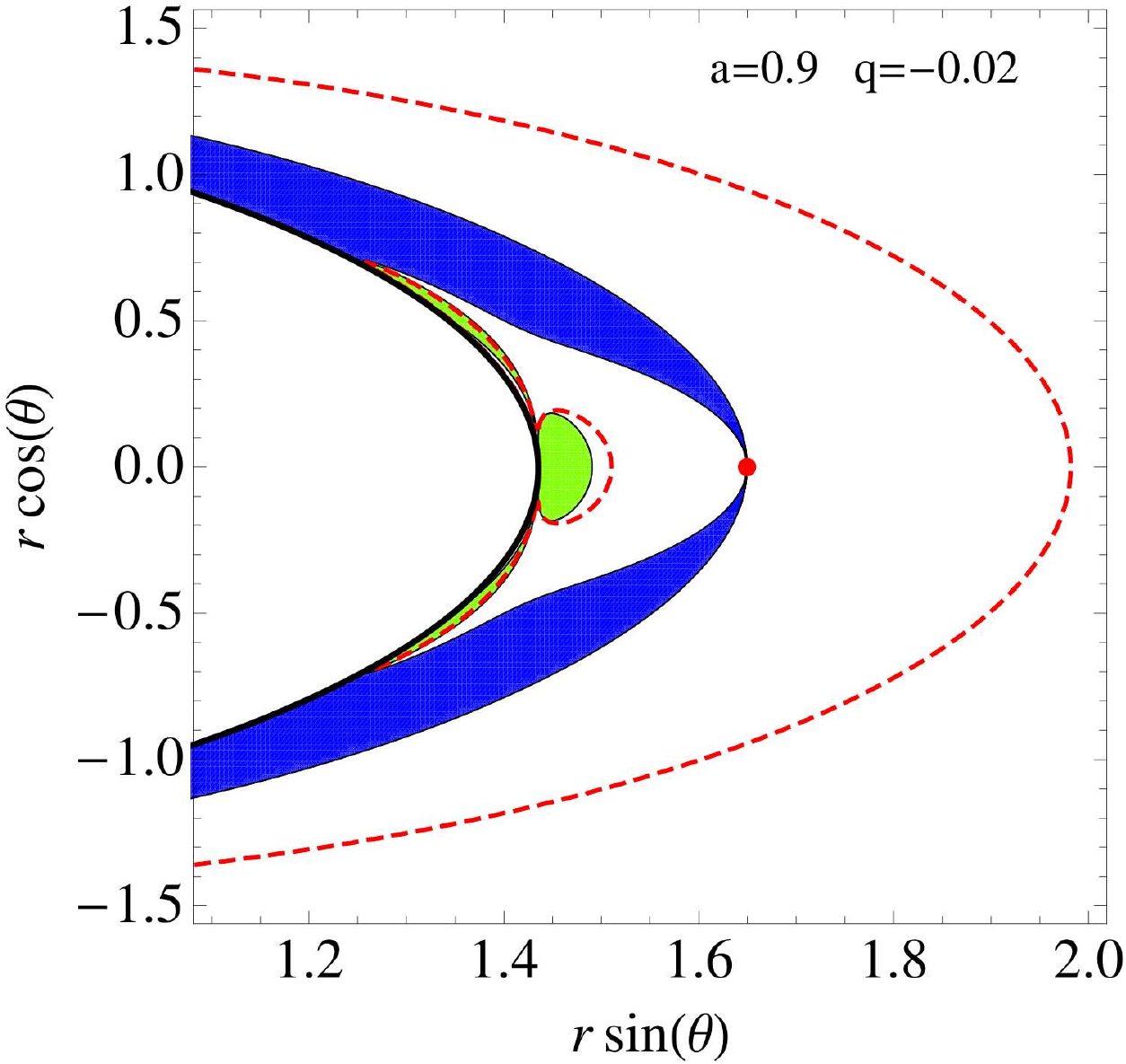} &
  \includegraphics[type=pdf,ext=.pdf,read=.pdf,width=0.32\textwidth]{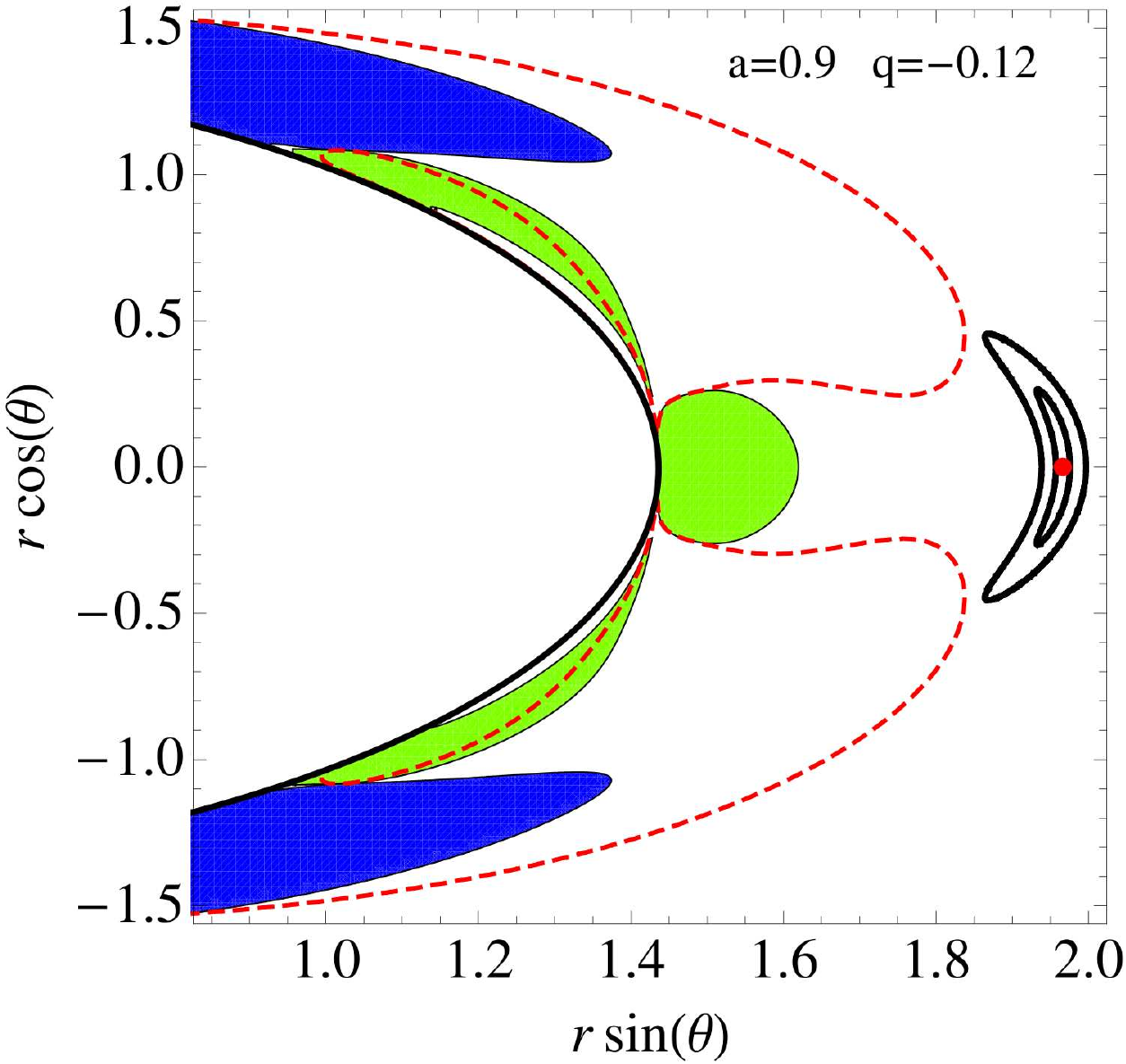} &
\includegraphics[type=pdf,ext=.pdf,read=.pdf,width=0.32\textwidth]{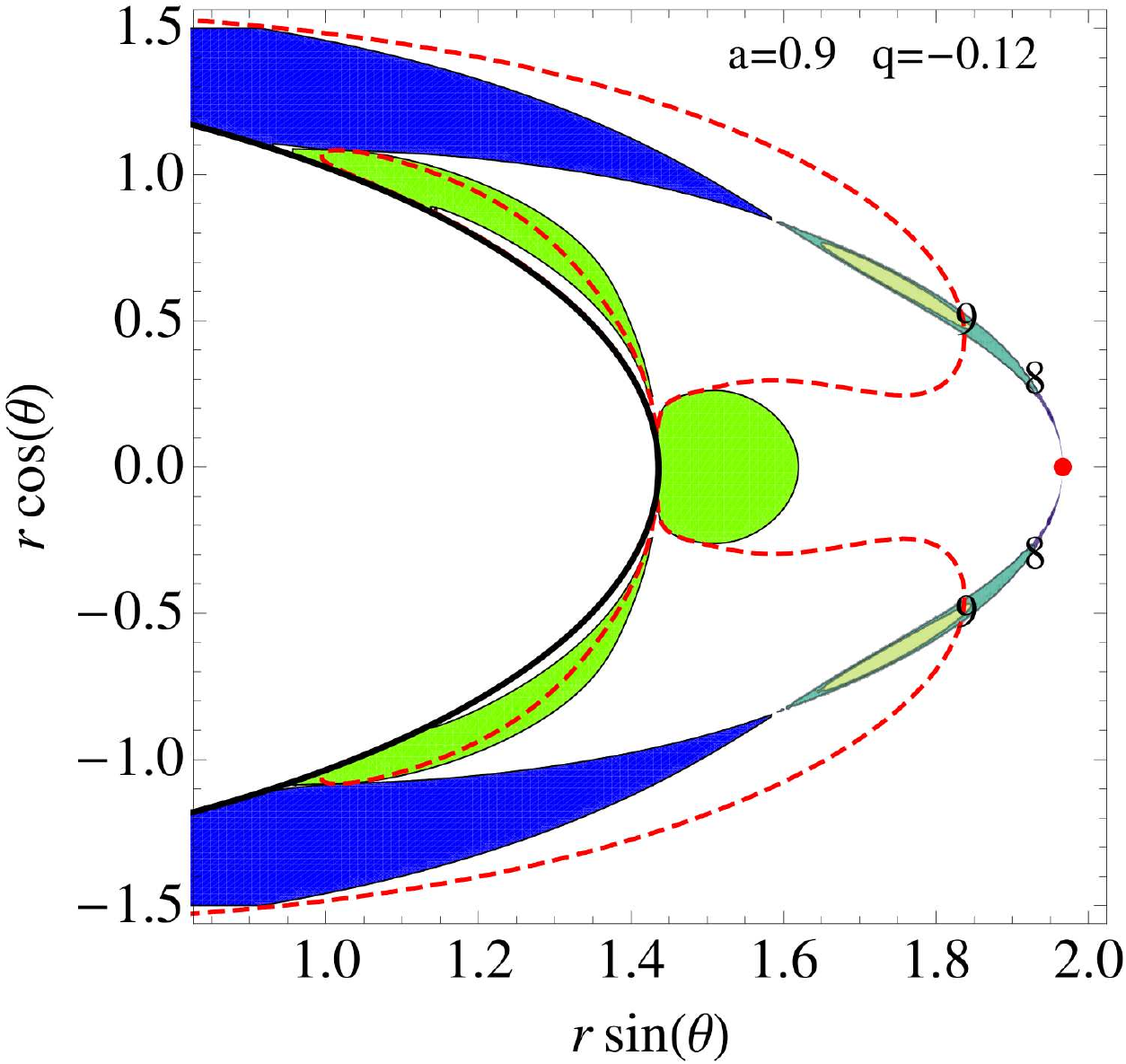} 
 \\
 \end{tabular}
 \\
 \end{tabular}
\caption{Same notation as in Fig.~\ref{fig_a_0.1}, but for $a = 0.9$ and $q=0.2$, $-0.015$, $-0.02$ and $-0.12$.
The accretion properties are qualitatively similar to those displayed for $a = 0.1$ in Fig.~\ref{fig_a_0.1}
and for $a = 0.5$ in Fig.~\ref{fig_a_0.5}, except for $q=-0.12$.
In that case, the ISCO is vertically unstable, and the gas does not plunge directly but remains
 trapped in the vicinities of the ISCO 
 (because $V_{\rm eff}(E_{\rm ISCO},L_{\rm ISCO},r,\theta)<0$ near the ISCO). However, the potential barrier is tall enough to allow the formation
of two thick disks, above and below the equatorial plane [scenario (2c)]. This contrasts with the cases $a=0.1$, $q=-4$ (Fig.~\ref{fig_a_0.1}) and 
$a=0.2$, $q=-2$ (Fig.~\ref{fig_a_0.5}), where the potential barrier is smaller and can be easily overcome 
if the gas is slightly perturbed.
\label{fig_a_0.9}}
\end{figure*}

Examples of the scenario outlined above are shown in Figs.~\ref{fig_a_0.1} (for $a=0.1$), 
\ref{fig_a_0.5} (for $a=0.5$) and \ref{fig_a_0.9} (for $a=0.9$), for various values of the anomalous quadrupole
moment $q$.
The solid black line at small radii is the partial horizon $x=1$,  the region where closed timelike curves exist
(i.e. the region where $g_{\phi\phi} < 0$) is shown in green/light gray, 
while the red dashed line is the boundary of the ergoregion (i.e. $g_{tt}=0$ on that line).
The position of the ISCO is marked by a red dot on the equatorial plane.
The thick disk, if it forms, sheds from the ISCO and 
is denoted by concentric contours, whose label is $\log_{\rm 10} T$,
where $T$ is the gas temperature (in K) assuming a polytropic equation of state $p=\kappa\rho_0^\Gamma$ ($\rho_0$ being the rest mass density) 
with $\Gamma=2$.
As we will show in Appendix~\ref{app}, the temperature for other values of the polytropic index $\Gamma$ can be obtained
by the simple rescaling $T(\Gamma)= T(\Gamma=2)\times 2 (\Gamma-1)/\Gamma$. Because the factor 
$2 (\Gamma-1)/\Gamma$ varies between $0$ and $1$
when $\Gamma$ varies in its allowed range $1<\Gamma<2$, the temperature plotted in our figures has to be interpreted as an upper value to
the real temperature.
In  blue/dark grey is  
the ``plunge'' region accessible to the gas shedding from the inner edges
of the thick disk [i.e. the region where $V_{\rm eff} (E_{\rm inner}, L_{\rm inner}, r, \theta) \ge 0$]. If no thick disk
is present, in  blue/dark grey is  the ``plunge'' region accessible to the gas shedding from the inner edge of the thin disk 
 [i.e. the region where $V_{\rm eff}(E_{\rm ISCO},L_{\rm ISCO},r, \theta)\geq 0$]: if this ``plunge'' region does not contain 
a neighborhood of the ISCO [cf. scenario (2b) above], we also show (with solid black lines around the ISCO) the 
contours $V_{\rm eff}(E_{\rm ISCO},L_{\rm ISCO},r,\theta)=-10^{-4}$ and  $V_{\rm eff}(E_{\rm ISCO},L_{\rm ISCO},r,\theta,)=-10^{-3}$,
which represent the region where the gas reaching the ISCO can move if subject to a small perturbation 
(e.g. if imparted a small initial velocity).

As can be seen, scenarios (1a), (1b) and (2a) happen for all the three values of the spin parameter that we consider. In particular,
although a more detailed scan of the parameter plane $(a,q)$ would be needed to draw ultimate conclusions, it seems that scenario (1a)
generically takes place when $q>q_1(a)$, scenario (1b) when $q_2(a)<q<q_1(a)$, and scenario (2a) when $q<q_2(a)$. However, for 
$q<q_2(a)$ one can also have scenario (2b) -- which we see for $a=0.1$, $q=-4$ and $a=0.5$, $q=-2$ -- or scenario (2c) -- which we
see for $a=0.9$ and $q=-0.12$. We have not been able to identify a scenario (2c) for $a=0.1$ or $a=0.5$.
This is because for these values of the spin, the potential barrier preventing the gas shedding at the ISCO from plunging
above and below the equatorial plane is smaller than for $a=0.9$, and a small perturbation is sufficient for the gas to plunge, thus making
the formation of a stationary thick disk impossible or at least very difficult.
Mathematically, this means that when varying the exponent $\beta$ in Eq.~\eqref{eq-leq}, 
we were unable to find a value $\beta_{\rm crit}$
for which the thick disk has an inner shedding point (and therefore a stationary configuration).
Scenario (2c) is instead possible for $a=0.9$ because in that case the potential barrier is higher [cf. the contours  
$V_{\rm eff}(E_{\rm ISCO},L_{\rm ISCO},r,\theta)=-10^{-4}$ and  $V_{\rm eff}(E_{\rm ISCO},L_{\rm ISCO},r,\theta,)=-10^{-3}$
for $a=0.9$ and $q=-0.12$ with those for $a=0.1$, $q=-4$ or $a=0.5$, $q=-2$].

Finally, let us comment on where the ``surface'' of the compact object may be located. 
As already mentioned, the MN metric is an exact vacuum solution of the Einstein equations,
and describes the spacetime \textit{outside} a generic compact object made
of exotic matter, which 
determines the free parameters of the MN metric (e.g. the anomalous quadrupole moment $q$). This object should of course
cover the pathological features of the MN solutions, such as the closed timelike curves, the curvature
singularity at $x=1$, $y=0$ and the partial horizon $x=1$, $|y|\leq1$. Even if this were the case, however, this scenario would still present some difficulties.
First, such an exotic compact object would probably be subject to gravitational instabilities on a dynamical timescale, unless its radius
is large enough to cover the whole ergoregion. In fact, the so-called ergoregion instability is known to happen on a dynamical timescale
for objects whose exterior metric resembles that of Kerr BH, at least for spin parameters $a\lesssim 2$~\cite{ergo}. Second, 
if the compact object has a surface, the kinetic energy of the plunging material must eventually be emitted by the surface, while 
in the case of a BH it simply gets lost in the event horizon. This argument, coupled with the observation that  BH candidates are indeed
 dimmer than systems known to contain neutron stars, or at least
stars with a surface, is a strong indication of the existence of event horizons~\cite{horizonEvidence}. However, we should stress that one can in principle 
imagine models escaping this argument (e.g. a gravastar, where the radiation emitted by the surface is so redshifted that it becomes
observationally negligible~\cite{gravastar}).

Another possibility is that GR breaks down
near the naked curvature singularity and the closed timelike curves of the MN geometry. One can therefore
imagine a completely regular spacetime that is described by the MN metric away from the 
singularity and the closed timelike curves, and which does not present these pathological features any more (see Refs.~\cite{string}
for some explicit ideas in this direction). An advantage of this scenario is that the horizon of this ``regular'' MN
spacetime would allow these objects to have a dimmer luminosity, in agreement with the observations~\cite{horizonEvidence},
and it may even quench the ergoregion instability or make it happen on a non-dynamical timescale.

Of course, the MN spacetimes that we consider, irrespective of their actual physical significance, 
are also an ideal tool to set-up null experiments to test the Kerr BH hypothesis,
since they reduce exactly to Kerr BHs when $q=0$. Indeed, 
any experiment pointing
at a value of $q$ significantly different from $0$
would imply that the object under consideration is
either a compact object different from a BH within GR, or a BH or a compact object in
a gravity theory different from GR. In this sense, MN spacetimes can be used not
only to test the BH hypothesis but also to test the gravity theory itself~\cite{scott_null_exp}.

\begin{table*}
\begin{center}
\begin{tabular}{c c c c c c c c c c c c c}
\hline \\
$a$ & \hspace{.5cm} & $q$ & 
\hspace{.5cm} & $\beta_{\rm crit}$ & \hspace{.5cm} & 
$1 - E_{\rm ISCO}$ & \hspace{.5cm} & $1 - E_{\rm inner}$
& \hspace{.5cm} & $L_{\rm ISCO}/E_{\rm ISCO}$ & \hspace{.5cm} & 
$L_{\rm inner}/E_{\rm inner}$ \\ \\
\hline 
0.1 & & $-0.5$ & & 0.0187 & & 0.069 & & 0.075 & & 3.454 & & 3.426 \\
0.5 & & $-0.2$ & & 0.074 & & 0.093 & & 0.117 & & 3.057 & & 2.990   \\
0.9 & & $-0.015$ & & 0.048 & & 0.167 & & 0.195 & & 2.448 & & 2.418   \\
0.9 & & $-0.12$ & & 0.204 & & 0.310 & & 0.332 & & 1.970 & & 1.921   \\
\hline
\end{tabular}
\end{center}
\caption{The effect of the thick disk on the accretion efficiency and on the spin evolution of
the central object, for the cases considered in Figs.~\ref{fig_a_0.1}, \ref{fig_a_0.5} and~\ref{fig_a_0.9} that present a thick disk inside the ISCO.}
\label{tab}
\end{table*}

\section{Discussion \label{s-d}}

The scenario (1a) discussed in the previous section is, as we have mentioned,
very similar to the accretion process in a Kerr spacetime. At the ISCO, which
is radially unstable, the gas plunges into
the compact object, remaining roughly on the equatorial plane.
Because this plunge takes place in a dynamical time and with no significant dissipation,
the infalling gas is not expected to emit a significant amount of radiation,
and this is a crucial ingredient of the  Novikov-Thorne model.
Under this assumption, the accretion luminosity  is simply
$L_{\rm acc} = \eta \dot{M} c^2$,
where $\eta = 1 - E_{\rm ISCO}$ is the radiative efficiency.
The evolution of the spin parameter is then regulated by~\cite{bar}
\be\label{eq-spin}
\frac{d a}{d \ln M} = \frac{1}{M}
\frac{L_{\rm ISCO}}{E_{\rm ISCO}} 
- 2 a \, ,
\ee  
neglecting the small correction coming from the radiation 
emitted by the disk and captured by the compact object.

In the case of a Kerr background, the Novikov-Thorne
model seems to be confirmed by recent three-dimensional general relativistic magnetohydrodynamic
simulations~\cite{cfa} (see however Ref.~\cite{krolik} for a different conclusion), and can therefore be 
used to interpret the X-ray spectra of BH candidates
and estimate their
spin~\cite{spin}. 
It seems therefore plausible that
the Novikov-Thorne model should work also in MN spacetimes,
and in principle one can generalize the technique used to estimate the 
spin parameter of BH candidates to measure possible deviations from the
Kerr geometry (e.g. the anomalous quadrupole moment $q$)~\cite{continuum}. From Eq.~(\ref{eq-spin}), it also follows
that the accretion process onto a compact object 
more oblate than a Kerr BH can potentially spin the body up 
to $a > 1$~\cite{evo}.

Scenarios (2a) and (2b) are quite similar to scenario (1a). In these cases the
ISCO is vertically unstable,
so the gas plunges above and below the equatorial plane, but this infall still happens on
a dynamical time and with a negligible emission of radiation. This presumably 
means that the Novikov-Thorne model is a reasonable description of the thin disk and that
the spin of the central object evolves according to \eqref{eq-spin}.
Also, because the shape of the plunge
region draws the gas towards the rotation axis, outflows and jets might possibly form
in these scenarios if magnetic fields are present.

In scenarios (1b) and (2c), the assumption of no radiation
emitted inside the ISCO is not valid, because of the presence of the
thick disks that we discussed in the previous section. 
As a result, the radiative efficiency is not $\eta = 1 - E_{\rm ISCO}$ like in the Novikov-Thorne model, 
but $\eta = 1 - E_{\rm inner}$, while the spin evolution of the central object is still described by Eq.~(\ref{eq-spin}),
but with the ratio $L_{\rm ISCO}/E_{\rm ISCO}$ replaced by 
$L_{\rm inner}/E_{\rm inner}$. The values of these quantities are shown in Table~\ref{tab} for the cases
shown in Figs.~\ref{fig_a_0.1}, \ref{fig_a_0.5} and~\ref{fig_a_0.9}, and as can be seen the corrections due to the presence of the 
thick disks inside the ISCO are about $1-3$\% for the ratio $L_{\rm inner}/E_{\rm inner}$, and
about $10-25$\% when it comes to $\eta$.

Moreover, the spectrum of the thick
disks inside the ISCO is quite different from that of the thin disk.
The temperature of a thin disk scales like
$T \sim M^{-0.25}$, $M$ being the disk's mass, while the temperature of the thick disks
inside the ISCO is independent of $M$ (see Appendix~\ref{app} for details), like
in the case of Bondi accretion flows. As can be seen from  Figs.~\ref{fig_a_0.1}, \ref{fig_a_0.5} and~\ref{fig_a_0.9},
we have $T\sim 10^{9}-10^{10}$~K~$\approx 0.1-1$~MeV in the thick disks, and 
at such
high temperatures the most efficient cooling mechanism
is thermal bremsstrahlung, whose emission rate is briefly reviewed in Appendix~\ref{app2}.

In the case of a 10~$M_\odot$ compact object, for reasonable 
values of the gas density of the thick disks, e.g. 10$^{12}$ 
particles/cm$^3$, the disk is definitely optically thin. The radiation 
emitted by the thick disk thus scales as $M^3$, but the flux 
observable on the Earth is completely negligible: considering
an object at a distance of 1~kpc, the intensity of the $\gamma$-ray
spectrum around 
$0.1-1$~MeV is lower than 
$10^{-15}$~$\gamma$~cm$^{-2}$~s$^{-1}$.
For a $10^9$~$M_\odot$ compact object, the thick disk inside the 
ISCO is not necessarily optically thin, but if it is, the 
$\gamma$-ray spectrum of the object may include a bump around
$0.1-1$~MeV, with the characteristic shape of thermal bremsstrahlung
(spectrum almost constant till energies $\sim T$ and then exponentially 
suppressed). If the object is at 10~Mpc from us, the flux on the Earth 
could be around 1~$\gamma$~cm$^{-2}$~s$^{-1}$, which is potentially
observable. For instance, the so-called MeV-blazars show this
feature~\cite{blaz}. Unfortunately, so far the spectrum of active galactic nuclei (AGN) is too poorly
understood to say anything conclusive, but the presence of a thick
disk inside the ISCO of the supermassive objects in galactic nuclei
will likely be testable in the future.

Finally, let us comment that although the presence of thick disks inside the ISCO
may be used, at least in principle, as an observational signature of the existence of non-Kerr compact objects,
the scenarios (1b) and (2c) where these disks form seem to arise only in limited regions of the $(a,q)$ plane. For instance,
as we have mentioned in the previous section, scenario (1b) seems to be possible only in a narrow region at
the transition between scenarios (1a) and (2a), while scenario (2c) seems only to be possible for sufficiently high spins.

\section{Conclusions \label{s-c}}

The $5-20M_\odot$ compact objects in X-ray binary systems 
and the supermassive bodies in galactic nuclei are currently
thought to be the Kerr BHs predicted by GR. The 
study of the electromagnetic radiation emitted in the accretion 
process onto these objects can be used to investigate their actual nature
 and therefore test the Kerr BH paradigm~\cite{continuum,iron,agn}.
In this paper we have studied the final stages of accretion,
when the gas reaches the inner edge of the thin accretion disk, located at the ISCO, 
and plunges into the compact 
object.
We find that for non-Kerr compact objects this process is much more complicated than
in the case of Kerr BH. More specifically, depending on the spin $a$ and anomalous quadrupole moment $q$ of the compact
object we find essentially four possible scenarios:
\begin{enumerate}
\item The ISCO is {\it radially} unstable, and the gas plunges into the  
compact object remaining roughly on the equatorial plane and without emitting significant radiation. 
This is the same scenario as in the Kerr case.
\item The ISCO is {\it radially} unstable and the gas 
plunges, but does not reach the compact object. Instead, 
it gets trapped between the object and the ISCO,  
forming a thick disk with $T\lesssim 10^{10}$ K and emitting by thermal bremsstrahlung.
\item The ISCO is {\it vertically} unstable, and the gas plunges into the                                                                
compact object {\it outside} the equatorial plane and without emitting significant radiation.
\item The ISCO is {\it vertically} unstable and the gas 
plunges, but does not reach the compact object. Instead,
it gets trapped between the object and the ISCO and        
forms two thick disks, above and below the equatorial plane.
 These thick disks have $T\lesssim 10^{10}$ K and emit by thermal bremsstrahlung.
\end{enumerate}

While the second and the fourth of these scenarios
seem to happen only for objects more prolate than Kerr, and even in
that case only in a limited region of the parameter space $(a,q)$,
they are nevertheless possible and may be testable with future data.
Our results therefore show that care is needed 
when using the Novikov-Thorne model with objects that are different from Kerr BHs, as the assumption of negligible 
radiation emission inside the ISCO may not be correct.
In particular, an excess of emission due to the presence of a thick disk in the region inside the ISCO might 
bias measurements
of the spin of BH candidates towards high values. It may be interesting to
test this possibility with the spectra of high-spin BH candidates such
as  GRS~1915+105~\cite{1915}, when future 
more accurate measurements of the distance to this object will be available.
For the time being, however, despite the peculiar features
of the accretion process that we discovered in this paper, compact objects more prolate than Kerr 
BHs cannot be ruled out by astrophysical observations.


\begin{acknowledgments}
We would like to thank Luciano Rezzolla
for critically reading this manuscript 
and providing useful feedback. 
We are also grateful to Sergei Blinnikov for 
useful discussions and suggestions.
The work of C.B. was supported by World Premier International 
Research Center Initiative (WPI Initiative), MEXT, Japan, and 
by the JSPS Grant-in-Aid for Young Scientists (B) No.~22740147.
E.B. acknowledges support from NSF Grants PHY-0903631.
\end{acknowledgments}


\appendix

\section{The theory of non self-gravitating, 
stationary and axisymmetric fluid configurations \label{app}}

Here we briefly review the theory of non self-gravitating, 
stationary and axisymmetric fluid configurations in generic stationary and axisymmetric
background spacetimes. For more details, we refer the reader to Refs.~\cite{pol1,pol2}
(but see also Refs.~\cite{hydrodrag,font_daigne,zanotti_etal}).

We start by considering a perfect fluid with 
4-velocity $\boldsymbol{u}^{\rm fluid}$, which is described by the stress-energy tensor
\begin{align}
\label{stress-tensor}
T^{\mu\nu}&= (\rho+p)u_{\rm fluid}^\mu u_{\rm fluid}^\nu+p g^{\mu\nu}\nonumber
        \\&= \rho_0 h u_{\rm fluid}^\mu u_{\rm fluid}^\nu+p g^{\mu\nu} \, .
\end{align}
Here $p$, $\rho_0$, $\rho$ and $h\equiv (p+\rho)/\rho_0$ are the
pressure, rest-mass density, energy density and specific enthalpy. We also assume that the fluid is described by a polytropic
equation of state $p=\kappa\rho_0^\Gamma=\rho_0\varepsilon(\Gamma-1)$,
where $\varepsilon=\rho/\rho_0-1$ is the internal energy per unit
rest-mass, and $\kappa$ and $\Gamma$ are respectively the polytropic constant and polytropic
index. Because we are neglecting the self-gravity of the
fluid, $\boldsymbol{g}$ is  the background metric (i.e., in our case,
the MN metric).
We assume that the fluid moves on non-geodesic circular orbits with 4-velocity
\begin{align}
&\boldsymbol{u}^{\rm fluid}=A(r,\theta)\left[\frac{\partial}{\partial t}+
\Omega(r,\theta)\frac{\partial}{\partial \phi}\right]\nonumber\\
&=U(r,\theta)\left[-dt+\ell(r,\theta) d\phi\right]\;,\label{eq:u}
\end{align}
where $\Omega\equiv u_{\rm fluid}^\phi/u_{\rm fluid}^t$ is the angular
velocity, $A\equiv u_{\rm fluid}^t$ is called the redshift factor,
$U\equiv -u^{\rm fluid}_t$ is the energy per unit mass as measured at
infinity and $\ell \equiv -u^{\rm fluid}_\phi/u^{\rm fluid}_t$ is the
angular momentum per unit energy as measured at infinity. Note that $\ell$ is
conserved for a stationary and axisymmetric flow in a stationary and axisymmetric spacetime~\cite{conserved_ell}.
The angular momentum per unit energy and the angular
velocity are related by
\begin{equation}\label{eq:omega}
\Omega=-\frac{g_{t\phi}+g_{tt}\ell}{g_{\phi\phi}+g_{t\phi}\ell}\,,\quad
\ell=-\frac{g_{t\phi}+g_{\phi\phi}\Omega}{g_{tt}+g_{t\phi}\Omega}\;,
\end{equation}
while $\boldsymbol{u}^{\rm fluid}\cdot\boldsymbol{u}^{\rm fluid}=-1$ implies
\begin{gather}
U=\sqrt{\frac{g_{t\phi}^2-g_{tt}\,g_{\phi\phi}}{g_{tt}\ell^2+2 g_{t\phi}\ell+g_{\phi\phi}}}\label{eq:U}\;,\\
A=\sqrt{\frac{-1}{g_{tt}+2 g_{t\phi}\Omega+g_{\phi\phi}\Omega^2}}\;,\label{eq:A}\\
AU=\frac{1}{1-\Omega\ell}\label{eq:AU}\;.
\end{gather}

To calculate the structure of the fluid configuration, we use Euler's
equation 
\eq\label{eq:euler} 
a_{\rm fluid}^\mu=-\frac{(g^{\mu\nu}+u_{\rm fluid}^\mu u_{\rm
fluid}^\nu)\partial_\nu p}{p+\rho}\;, 
\eeq 
where $a_{\rm fluid}^\mu$
is the fluid's 4-acceleration. If the pressure is
assumed to depend only on $r$ and $\theta$ (which follows from the stationarity and axisymmetry of the fluid and geometry) 
and if the equation of
state is barotropic [\textit{i.e.,} if $\rho=\rho(p)$]\footnote{This
is the case for a polytropic equation of state, as
$\rho=p/(\Gamma-1)+(p/\kappa)^{1/\Gamma}$.}, 
one can express Eq.~\eqref{eq:euler} in terms of the
gradient of a scalar potential $W(p)$:
\eq\label{eq:W_def}
a^{\rm fluid}_\mu=\partial_\mu W \;, \quad W(p)= -\int^p
\frac{dp'}{p'+\rho(p')}\;. \eeq 
Also, from the definition
of 4-acceleration 
and using Eqs. \eqref{eq:u}, \eqref{eq:A}, \eqref{eq:AU} and the Killing
equations $\nabla_{(\mu}\xi_{\nu)}=0$ for $\xi=\partial /\partial t$
and $\xi=\partial /\partial \phi$, one obtains
\eq\label{eq:euler2} a^{\rm fluid}_\mu=\partial_\mu W= -\frac{\partial_\mu
p}{p+\rho}= \partial_\mu \ln
U-\frac{\Omega}{1-\Omega\ell}\partial_\mu\ell\;. 
\eeq

Taking the derivative of this equation,
anti-symmetrizing and using the trivial fact that
$\partial_{[\mu\nu]}W=\partial_{[\mu\nu]}\ell=\partial_{[\mu\nu]}U=0$,
we obtain $\partial_{[\mu}\Omega\,\partial_{\nu]}\ell=0$. This
implies $\nabla\Omega\propto\nabla\ell$, and therefore $\ell$ and
$\Omega$ have the same contour levels~\cite{vonzeipel}. One can then express $\Omega$ as a function of $\ell$, 
$\Omega=\Omega(\ell)$, and write
Eq. \eqref{eq:euler2} in the integral form
\begin{align}
&W-W_{\rm out}= -\int_0^p \frac{dp'}{p'+\rho(p')}\nonumber\\& =\ln U
-\ln U_{\rm out} -\int_{\ell_{\rm
out}}^{\ell}\frac{\Omega(\ell')d\ell'}{1-\Omega(\ell')\ell'}\;,
\label{eq:euler_int}
\end{align}
where $W_{\rm out}$ and $\ell_{\rm out}$ are the potential and
angular momentum per unit energy at the outer edge of the
fluid configuration. (Of course, $W_{\rm out}$ and $\ell_{\rm out}$ can be replaced
by the values of $W$ and $\ell$ at any other point of the fluid's boundary, e.g. 
the inner edge of the configuration if
that is present.) 

In practice, in order to calculate $W-W_{\rm out}$, 
one specifies the functional form of $\ell$ on the equatorial plane, $\bar{\ell}(r)\equiv\ell(r,\theta=\pi/2)$. From
that, one can use Eq.~\eqref{eq:omega} to obtain the angular velocity on the equatorial plane,  
$\bar{\Omega}(r)\equiv\Omega(r,\theta=\pi/2)$, 
and calculate the integral in Eq.~\eqref{eq:euler_int} as
\begin{equation}
\int \frac{\Omega(\ell')d\ell'}{1-\Omega(\ell')\ell'}=
\int \frac{\bar{\Omega}(r')}{1-\bar{\Omega}(r')\bar{\ell}(r')} \frac{d\bar{\ell}(r')}{dr'} dr'\,.
\end{equation}
To calculate $\ln U (r,\theta)$, one needs instead to reconstruct the functional form of $\ell(r,\theta)$ outside the equatorial plane. 
This can be done by solving $\Omega(r,\theta)=\bar{\Omega}(\bar{r})$ for the equatorial radius $\bar{r}$
where the angular velocity has the same value as at a given point $(r,\theta)$ outside the equatorial plane. Using Eq.~\eqref{eq:omega}
and the fact that $\Omega(r,\theta)=\bar{\Omega}(\bar{r})$ implies   $\ell(r,\theta)=\bar{\ell}(\bar{r})$ 
(because  $\ell$ and $\Omega$ have the same contour levels),
this equation becomes
\eq
\frac{g_{t\phi}(r,\theta)+g_{tt}(r,\theta)\bar{\ell}(\bar{r})}{g_{\phi\phi}(r,\theta)+g_{t\phi}(r,\theta)\bar{\ell}(\bar{r})}=
\frac{\bar{g}_{t\phi}(\bar{r})+\bar{g}_{tt}(\bar{r})\bar{\ell}(\bar{r})}{\bar{g}_{\phi\phi}(\bar{r})+\bar{g}_{t\phi}(\bar{r})\bar{\ell}(\bar{r})}\,,
\eeq
where we have denoted with $\bar{g}_{\mu\nu}$ the metric on the equatorial plane. This equation
 is known as the equation of the von Zeipel cylinders, and by allowing one to express $\bar{r}$ as a function of $r$ and $\theta$,
it permits calculating both $\Omega(r,\theta)$ and $\ell(r,\theta)$ outside the equatorial plane.

The integral Euler equation \eqref{eq:euler_int} can be
further simplified if the equation of state is
polytropic, because in this case 
\eq\label{eq:euler_int3} \int_0^p
\frac{dp'}{p'+\rho(p')}=\ln\frac{h}{h_{\rm out}} \,,\eeq
where $h_{\rm out}$
is the specific enthalpy at the outer edge of the fluid configuration. Because for
a polytropic equation of state the enthalpy is
\begin{equation}
h = 1+\frac{\Gamma}{\Gamma-1}\kappa \rho_0^{\Gamma-1}\,,
\label{eq:h}
\end{equation}
one has $h_{\rm out}=1$ (because $p=\rho_0=0$ at the outer
edge of the fluid configuration), and Eqs. \eqref{eq:euler_int}
and \eqref{eq:euler_int3} give 
\eq \rho_0(r,\theta) =
\left\{\frac{\Gamma-1}{\Gamma}\frac{\left[e^{W_{\rm
out}-W(r,\theta)}-1\right]} {\kappa}\right\}^{1/(\Gamma-1)}\;,
\label{eq:rho} \eeq 
which makes it clear that the fluid occupies the region where $W(r,\theta)\leq W_{\rm out}$.

Once the rest-mass distribution is known, the
total rest mass is
\begin{equation}
M_{t,0} = \int\rho_0\sqrt{-g}u_{\rm fluid}^t d^3x \ ,
\end{equation}
where $d^3 x=dr \, d\theta\, d\phi$
is the coordinate 3-volume element, while the mass-energy is
\begin{equation}
M_{\rm t} = \int (T^r_r + T^\phi_\phi +
        T^\theta_\theta - T^t_t)\sqrt{-g}\,d^3x\,.
\end{equation}

From Eq.~\eqref{eq:rho} one can also calculate the temperature using $p=n k_{\rm B} T$ (where $k_{\rm B}$
is the Boltzmann constant and $n$ is the number density). Using the polytropic equation of state, one gets
\eq
T=\frac{m_{\rm p} \kappa \rho_0^{\Gamma-1}}{k_{\rm B}}=\frac{2(\Gamma-1)}{\Gamma}\times\frac{m_{\rm p}\left[e^{W_{\rm
out}-W(r,\theta)}-1\right]}{2 k_{\rm B}} \,,
\eeq
where we have assumed that the mass of the gas particles is given by the proton mass $m_{\rm p}$. We notice that 
the temperature is independent of the mass of the fluid configuration, and that
the factor $2(\Gamma-1)/{\Gamma}$ varies between 0 and 1 when $\Gamma$ varies in its allowed range $1<\Gamma<2$.

\section{Thermal bremsstrahlung \label{app2}}

The emission rate per unit volume for a fully ionized hydrogen gas
due to bremsstrahlung processes is~\cite{svensson}
\eq
\Lambda_{brems} = \alpha r_e^2 m_e c^3 n^2 F(\Theta_e) \, ,
\eeq
where $\alpha = 1/137$ is the electromagnetic fine structure constant,
$r_e$ is the classical electron radius, $m_e$ is the electron mass, $c$
is the speed of light, $n$ is the number density of electrons (protons),
and $\Theta_e = k_B T_e/m_2 c^2$ is the dimensionless electron
temperature. Here $F$ is the dimensionless radiation rate due to
relativistic bremsstrahlung. It is convenient to write
$F = F_{ee} + F_{ep}$, where $F_{ee}$ is the contribution from 
electron-electron collisions, while $F_{ep}$ the one from electron-proton
collisions. For $\Theta_e \le 1$, the approximated formulas are
\be
F_{ee} &=& \frac{20}{9 \sqrt{\pi}} (44 - 3 \pi^2)
\Theta_e^{1.5} (1 + 1.1 \Theta_e + \Theta_e^2 - 1.25 \Theta_e^{2.5}) 
\nonumber\\
F_{ep} &=& \frac{32}{3} \sqrt{\frac{2}{\pi}}
\Theta_e^{0.5 }(1 + 1.78 \Theta_e^{1.34}) \, .
\ee
For $\Theta_e \ge 1$, we have instead
\be
F_{ee} &=& 24 \Theta_e \left[ \ln(2 \eta_E \Theta_e) + 1.25 \right] \, ,
\nonumber\\
F_{ep} &=& 12 \Theta_e \left[ \ln(2 \eta_E \Theta_e + 0.42) + 1.5 \right] \, ,
\ee
where
$\eta_E = \exp(-\gamma_E)$ and $\gamma_E \approx 0.5772$ 
is the Euler's number.


\end{document}